\documentclass[twocolumn]{aastex62}

\usepackage{longtable}

\usepackage{natbib}
\usepackage{graphicx}
\newcolumntype{H}{>{\setbox0=\hbox\bgroup}c<{\egroup}@{}}

\graphicspath{{./}{figures/}}

\received{January 1, 2019}
\revised{January 7, 2019}
\accepted{January 8, 2019}

\submitjournal{ApJS}
\shorttitle{Hildas with K2}
\shortauthors{Szab\'o et al.}
\begin{document}

\title{Rotational properties of Hilda asteroids observed by the K2 mission}

\correspondingauthor{Gyula Szab\'o}
\email{szgy@gothard.hu}

\author{Gyula M. Szab\'o}
\affiliation{ELTE E\"otv\"os Lor\'and University, Gothard Astrophysical Observatory, Szombathely, Hungary}
\affiliation{MTA-ELTE Exoplanet Research Group, 9700 Szombathely, Szent Imre h. u. 112, Hungary}

\author{Csaba Kiss}
\affiliation{Konkoly Observatory, Research Centre for Astronomy and Earth Sciences, \\ Konkoly Thege Mikl\'os \'ut 15-17, H-1121 Budapest, Hungary}

\author{R\'obert Szak\'ats}
\affiliation{Konkoly Observatory, Research Centre for Astronomy and Earth Sciences, \\ Konkoly Thege Mikl\'os \'ut 15-17, H-1121 Budapest, Hungary}

\author{Andr\'as P\'al}
\affiliation{Konkoly Observatory, Research Centre for Astronomy and Earth Sciences, \\ Konkoly Thege Mikl\'os \'ut 15-17, H-1121 Budapest, Hungary}
\affiliation{E\"otv\"os Lor\'and University, P\'azm\'any P\'eter s\'et\'any 1/A, H-1171 Budapest, Hungary}

\author{L\'aszl\'o Moln\'ar}
\affiliation{Konkoly Observatory, Research Centre for Astronomy and Earth Sciences, \\ Konkoly Thege Mikl\'os \'ut 15-17, H-1121 Budapest, Hungary}
\affiliation{MTA CSFK Lend\"ulet Near-Field Cosmology Research Group}

\author{Kriszti\'an S\'arneczky}
\affiliation{Konkoly Observatory, Research Centre for Astronomy and Earth Sciences, \\  Konkoly Thege Mikl\'os \'ut 15-17, H-1121 Budapest, Hungary}

\author{J\'ozsef Vink\'o}
\affiliation{Konkoly Observatory, Research Centre for Astronomy and Earth Sciences, \\ Konkoly Thege Mikl\'os \'ut 15-17, H-1121 Budapest, Hungary}

\author{R\'obert Szab\'o}
\affiliation{Konkoly Observatory, Research Centre for Astronomy and Earth Sciences, \\ Konkoly Thege Mikl\'os \'ut 15-17, H-1121 Budapest, Hungary}
\affiliation{MTA CSFK Lend\"ulet Near-Field Cosmology Research Group}


\author{G\'abor Marton}
\affiliation{Konkoly Observatory, Research Centre for Astronomy and Earth Sciences, \\ Konkoly Thege Mikl\'os \'ut 15-17, H-1121 Budapest, Hungary}

\author{L\'aszl\'o L. Kiss}
\affiliation{Konkoly Observatory, Research Centre for Astronomy and Earth Sciences, \\  Konkoly Thege Mikl\'os \'ut 15-17, H-1121 Budapest, Hungary}
\affiliation{Sydney Institute for Astronomy, School of Physics A29, University of Sydney, NSW 2006, Australia}


\begin{abstract}
Hilda asteroids orbit at the outer edge, or just outside of the Main Belt, occupying the 2:3 mean motion resonance with Jupiter. It is known that the group shows a mixed taxonomy that suggests the mixed origin of Hilda members, having migrated to the current orbit both from the outer Main Belt and from the Trojans swarms. But there are still few observations for comparative studies that help in understanding the Hilda group in deeper details. We identified 125 individual light curves of Hilda asteroids observed by the K2 mission. We found that despite of the mixed taxonomies, the Hilda group highly resembles to the Trojans in the distribution of rotation periods and amplitudes, and even the LR group (mostly C and X-type) Hildas follow this rule. Contrary to the Main Belt, Hilda group lacks the very fast rotators. The ratio  of extremely slow rotators ($P>$100~h) is a surprising 18\%{}, which is unique in the Solar System. The occurrence rate of asteroids with multiple periods (4\%{}) and asteroids with three maxima in the light curves (5\%{}) can be signs of high rate of binarity, which we can estimate as 25\% within the Hilda group.
\end{abstract}


\section{Introduction} 

Hilda asteroids occupy the region between the outer Main Belt (MB) and the Jupiter Trojan swarms (Trojans hereafter), in 3:2 mean motion resonance with Jupiter. Due to the dynamical stability, the group is well defined in the proper element space, and two collisional families, {Schubart} and Hilda, can be confirmed around mean inclinations of 3 and 9 degrees, respectively \citep{1982CeMec..28..189S, 2008MNRAS.390..715B, 2015MNRAS.454.2436V}. A recent estimate by \citet{2018AJ....156...30T} suggests the existence of $\sim 10$ thousand Hildas larger than 2 km, and a size distribution index of $\alpha = 0.38 \pm 0.02$. Extrapolating this power-law distribution to the 1 km range, we get an order of magnitude estimate of $\sim 10^5$ Hildas larger than 1 km, which represents a few percent of the $\sim 2$ million asteroids expected in the MB, as well as the 1--2 million asteroids larger than 1 km in the Trojan swarms. Thus, the Hilda group is 1--1.5 orders of magnitude less populated than the Trojans or the entire Main Belt, which still makes Hildas significant contributors to the Solar System small bodies, forming a bridge between the MB and the Trojan swarms.

For Hildas, the effect of perturbations from Jupiter are amplified by the 3:2 mean motion resonance, leading to a significant variation of the osculating elements. The group is named after (153) Hilda, its first discovered member. In the rotating reference system of Jupiter's orbit, a typical Hilda orbits along a ``Hilda triangle'', drawing a loop and residing in the triaxial libration points in aphelion for significant amount of time, while close to perihelion it transits to the consecutive libration point much faster. This way three density waves are formed along the orbit of Hildas, and two of these overdense regions orbit 60 degrees before and after Jupiter's actual longitude. There is a significant overlap between the Hildas and the Trojans, but due to the stability of the different families, the rate of exchanged objects is debated \citep{2018AJ....156...30T}. There are also interactions between Hildas and the outer MB, since Hildas immerse in the MB outskirts close to their perihelion. So, due to their position and their motion, it is plausible that both MB asteroids and Trojan asteroids could contribute to the Hilda group members.

The power index of differential size distribution ($\alpha$) is known to gradually decrease in the MB families (from 0.76 to 0.56 at the bright end, from 0.46 to 0.40 at the faint end), and lacking a definite pattern in the MB background \citep{2008Icar..198..138P}. $\alpha \approx 0.38$ \citep{2018AJ....156...30T} of Hilda asteroids fits nicely into the power index distribution of the MB. However, Trojan asteroids again follow a steeper size distribution, around $\alpha=0.44$ \citep{2007MNRAS.377.1393S} with no significant break point, which reflects a separate dynamical evolution of the Trojan swarms and the Main Belt.

\begin{figure}
\includegraphics[viewport=17 134 440 355,width=8cm]{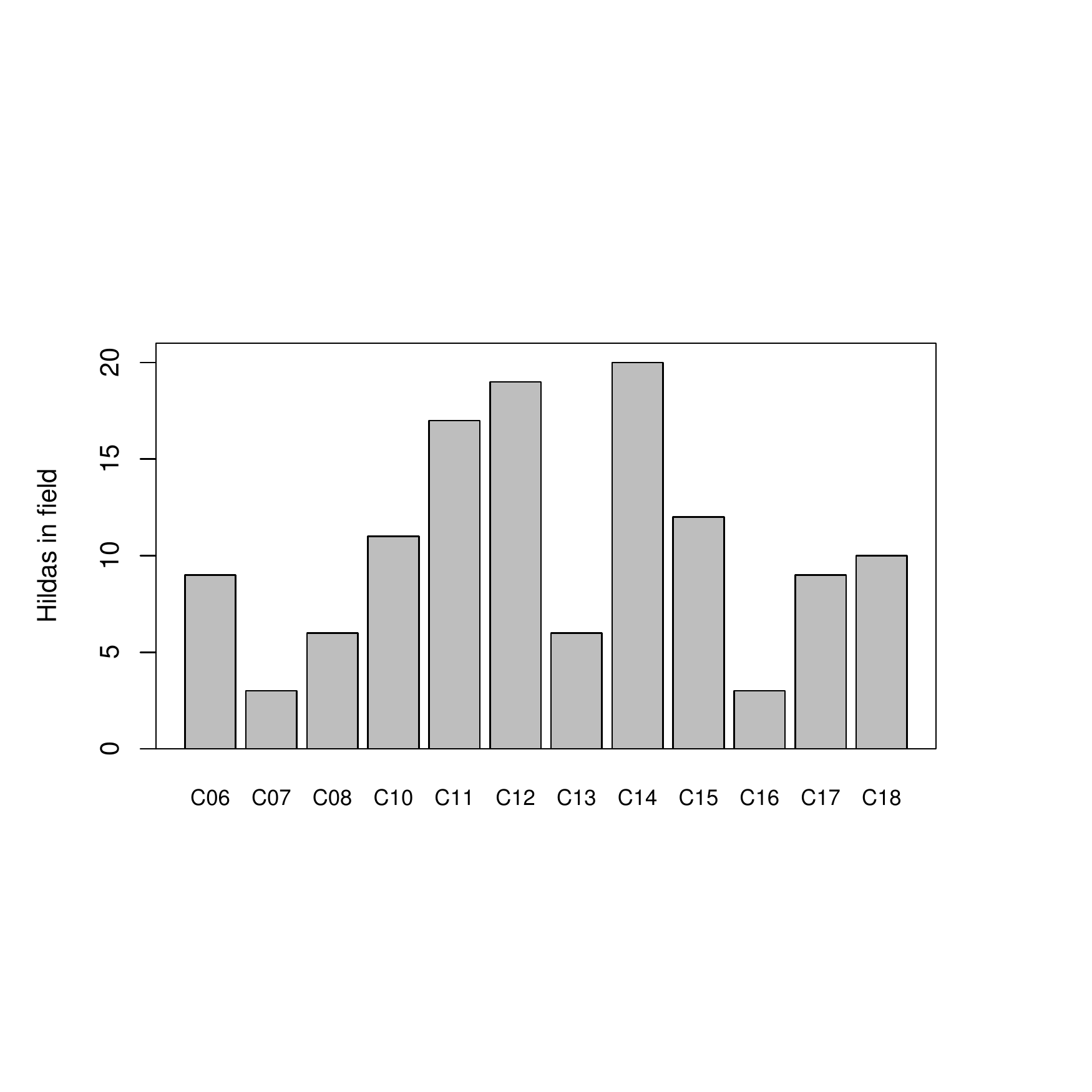}
\caption{The distribution of Hilda detections in K2 fields. Multiply detected asteroids are multiply counted.}
\label{fig:detections}
\end{figure}

\begin{figure}
\includegraphics[viewport=15 20 480 450,width=8cm]{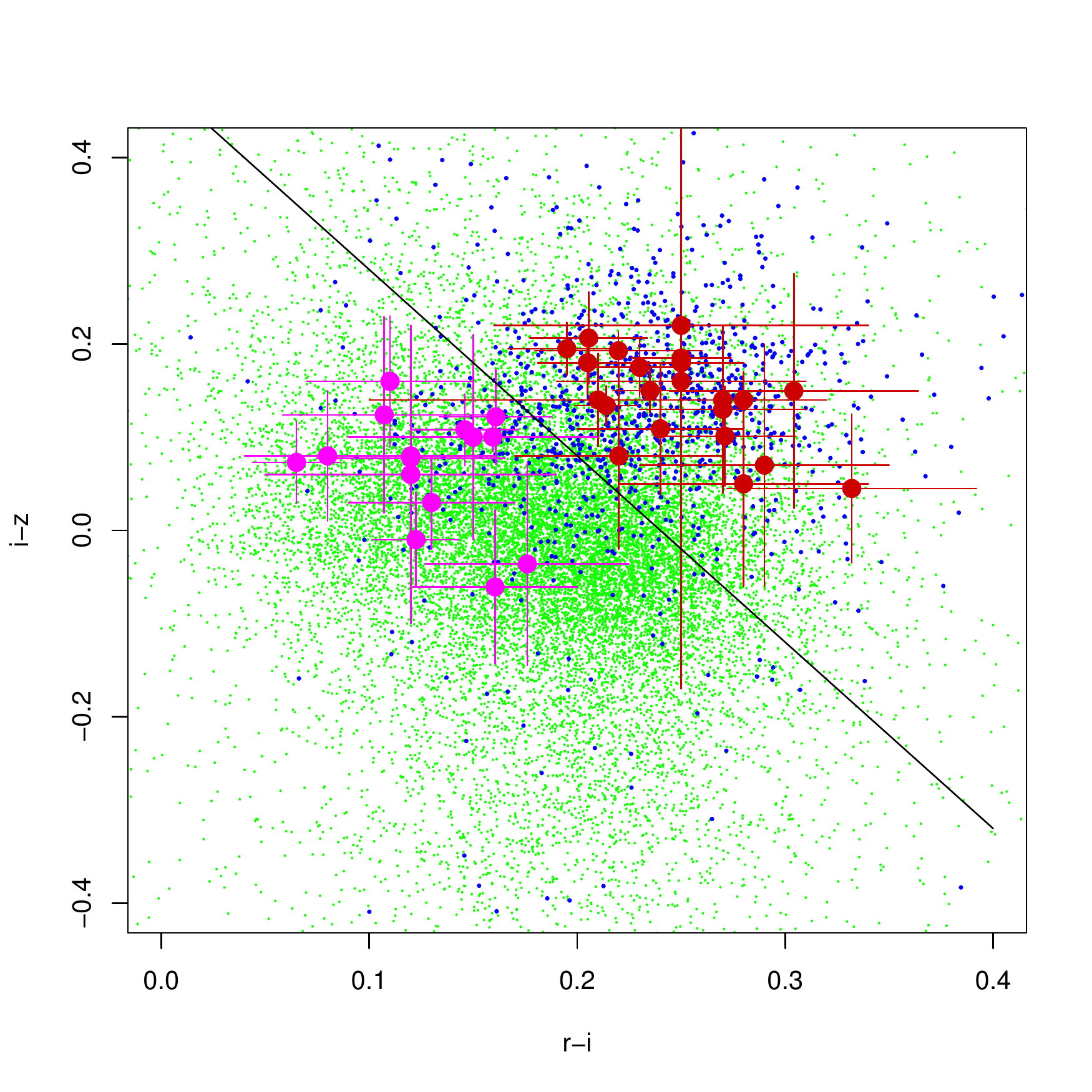}
\caption{The distribution of Hilda asteroids, detected by SDSS, on the $(r-i)$ - $(i-z)$ color-color diagram. The meaning of the colors are as follows: green: all Main Belt asteroids in SDSS MOC; blue: known Trojans in SDSS MOC; magenta: less red (LR) Hildas observed by K2 and with an entry in SDSS; red: red (R) Hildas observed by K2 and SDSS.}
\label{fig:sdss-colors}
\end{figure}

It is suggested that Trojans and outer MB asteroids exhibit different taxonomies (mostly C and X in the outer MB, mostly D and P in the Trojan swarms). Consistently with the dynamical position, D and P type Trojan asteroids, and C and X type outer MB asteroids are also observed in the Hilda group. \citet{2014AJ....148..112W} suggested a terminology of Red and Less Red Hildas (R and LR, respectively). \citet{2013DPS....4520505D} defined selection criteria for different taxonomy classes in the $gri$ slope vs.\ $i-z$ parameter space. According to their criteria, there are C and X asteroids in the LR group, and R Hildas consist mostly P and D types. 

Early solar system formation theories stated that Hildas originated in the middle solar system and were captured into their present-day orbits during a period of smooth migration \citep[e.g.,][]{Franklin2004}. Current solar system evolution models, however, mostly agree in a scenario in which the gas giants crossed a mutual mean-motion resonance sometime after the era of planet formation, resulting in a notable dynamical restructuring \citep[e.g.,][]{2005Natur.435..459T,2012AJ....144..117N}. According to these models many planetesimals which formed in the outer solar system were scattered inward during this period of dynamical instability; in this framework present-day Hildas and Jovian Trojans originate almost exclusively in these populations \citep[][]{Morbidelli2005,Roig2015}. 

\citet{2008Icar..193..567G,2008A&A...483..911R,2013A&A...549A..34A,2017AJ....153...69W} and \citet{2018Icar..311...35D} studied the color distributions of Hildas and Jovian Trojans and they found them to be consistent with a scenario in which the color bimodality in both populations developed before they were implanted into their present-day orbits. They propose that the shallower magnitude distribution of the Hildas is a result of an initially much larger Hilda population, which was subsequently depleted as smaller bodies were ejected from the narrow 3:2 resonance via collisions, also suggesting a common origin for Hildas and Trojans as predicted by current dynamical instability theories of solar system evolution.

\citet{2017AJ....153...69W} investigated the near infrared spectra of Hildas and found that Trojans and Hildas possess similar overall spectral shapes, suggesting that the two minor body populations share a common progenitor population. A more detailed examination reveals that while the red Trojans and Hildas have nearly identical spectra, less-red Hildas are systematically bluer in the visible and redder in the near-infrared than less-red Trojans.
They argue that the less-red and red objects found in both Hildas and Trojans represent two distinct surface chemistries and attribute the small discrepancy between less-red Hildas and Trojans to the difference in surface temperatures between the two regions.


Trojans cover a wider range of inclination than the Main Belt, many members going up to 30 degrees inclination, while the eccentricity range of Hildas covers half of that of Trojans, and up to 20 degrees inclination represented by a few members. In essence, Hildas fit more into the Main Belt families in terms of size distribution characteristics; in position their orbits overlap with the Trojan swarms; and  taxonomy suggests a region of mixed material of the Trojan swarms and the outer MB.

 The comparison of the Trojan and the MB observations shows that there are significant differences between the rotation properties of MBs and Trojans, the average rotation period is lower and the rate of binaries is very high in the Trojan swarms \citep{2017A&A...599A..44S}. Also, distribution of the minimum density is truncated at a lower value for Trojan asteroids, which is an evidence of significant porosity, in comparison with the MB asteroids. Since the Hilda group is suspected to share the taxonomies of the outer MB and the Trojans, and these large asteroid reservoirs are dynamically coupled via the Hilda group, it is plausible that Hilda asteroids also exhibit a mix of rotation properties of the MB and the Trojan families. It seemed also plausible that the specific rotation properties characteristic to the MB and Trojan asteroids are also preserved in the LR and R group of Hildas, respectively. We investigate this question in this paper, and disproof this simple belief.

The scope of this paper is to provide unbiased period and amplitude distributions of Hilda asteroids, and also, to test differences between the rotational properties of Hildas of Main Belt-like and Trojan-like taxonomy.

\section{Observations}

The K2 mission observed more than a hundred Hilda asteroids, enabling a detailed analysis of this group. More specifically, trails of 103 Hildas were detected in the K2 Campaigns 6--18. There are multiple detections of 22 Hildas, and in total, there are 125 observations. We did not recover data for 15 objects that were observed in the very crowded stellar fields of Campaigns 7 and 11. The distribution of detected Hilda asteroids is plotted in Fig.~\ref{fig:detections}.

The log of observations is shown in the Appendix (Table 4), listing all detections, and providing separate lines in the case of multiple observations. This is necessary since the  observational geometry could have changed significantly, also leading to e.g. the change of light curve amplitude due to the difference aspect angle.  

The light curves were extracted by our tools developed to obtain photometry of moving objects in the K2 fields, following the processing scheme developed in \citet{2015ApJ...804L..45P, 2016AJ....151..117P}, \citet{2016MNRAS.457.2908K}, and \citet{2018ApJS..234...37M}. 
The pipeline is based on the FITSH software package \citep{fitsh}. We registered the frames to the same reference system in order to perform differential image analysis. Astrometric solutions were derived for every mosaic frames taken during the campaign using the Full Frame Images (acquired once per campaign) as templates, to register the individual frames. In some cases we also enlarged the images by $\sim 3$ times and transformed them into RA-Dec directions. This subpixel-level re-sampling and spatial transformation helped to decrease the fringing of the residual images in the next step, and, more importantly, allowed us to extract usable light curves for a few targets where the Point Spread Function (PSF) was not fully covered by the pre-selected pixel masks. We then subtracted a median image from all frames in the following way: a series of frames were drawn from the full sample to form a master median-combined image, which was then subtracted from the subsequent frames. The median frame was created from a subset of frames that did not contain the target. We applied simple aperture photometry to the differential images based on the ephemeris provided by the JPL HORIZONS service \citep{horizons}. 

The light curves obtained were analysed with a residual minimization algorithm \citep{2016AJ....151..117P, 2018ApJS..234...37M}.
In this method we fit the data with a function 
\begin{equation}
f(t) ~=~ A + B \cos(2\pi f \Delta t) + C \sin(2\pi f \Delta t),
\end{equation}
where $f$ is the trial frequency, $\Delta t = T - t$,  $T$ the approximate center of the time series, and A, B, and C are fit parameters to be determined. After folding the data with the trial frequency $f$, the dispersion of the folded light curve is calculated in $N \sim 10$ frequency bins as a function of $f$.   
We search for the minima of the dispersion curves for each frequency. As it is demonstrated in \citet{2018ApJS..234...37M}, the best-fit frequencies obtained with this method are identical to the results of LombScargle periodogram or fast Fourier transform analyses, with a notably smaller general uncertainty in the residuals. 

A large fraction of K2 Hilda asteroids were also detected in the Sloan Digital Sky Survey (SDSS). Half of the detections had an entry field in the SDSS Moving Object Catalog \citep[MOC;][]{2001AJ....122.2749I,2002AJ....124.1776J,2008Icar..198..138P}, while the other detections were listed in SDSS PhotoObj files, not recognized as moving targets. The detections were identified using the Solar System Object Image Search facility provided by the Canadian Astronomy Data Centre\footnote{ https://www.cadc-ccda.hia-iha.nrc-cnrc.gc.ca/en/ssois/}. After collecting all K2 Hilda detections from SDSS MOC and PhotoObj files, we plotted the color distributions in the $(r-i)$--$(i-z)$ color-color space (Fig.~\ref{fig:sdss-colors}). It is seen that Red and Less Red Hildas are convincingly separated (although some asteroids have error bars crossing the suggested boundary).

The measured rotation periods and amplitudes of all Hilda asteroids, and comparison to previous data in the literature is shown in the Appendix (Table 5). The complete collection of light curves, folded light curves and time-frequency distibutions are also shown for all K2 Hilda asteroids in the Appendix.


\begin{figure}
\includegraphics[width=\columnwidth]{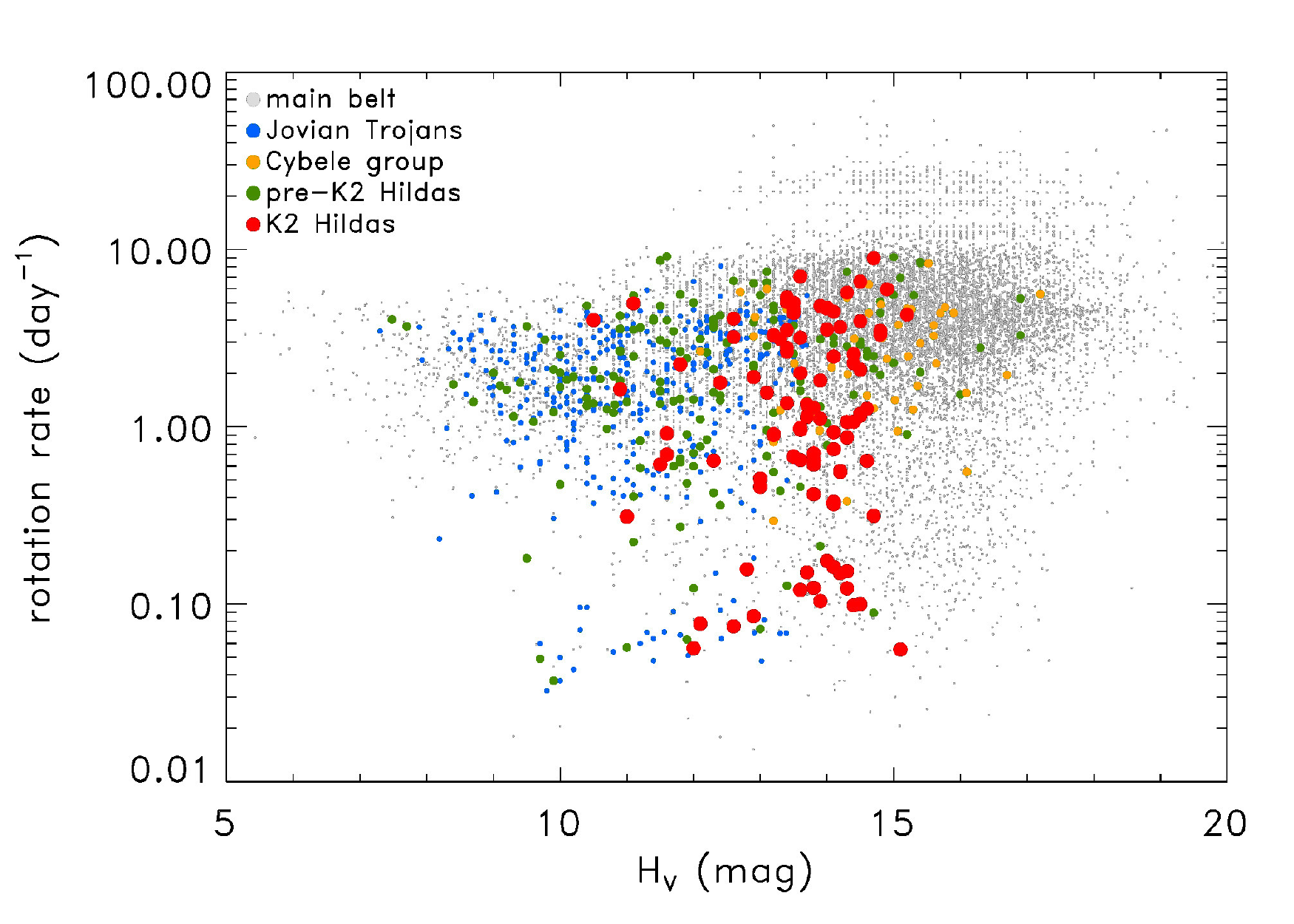}
\includegraphics[width=\columnwidth]{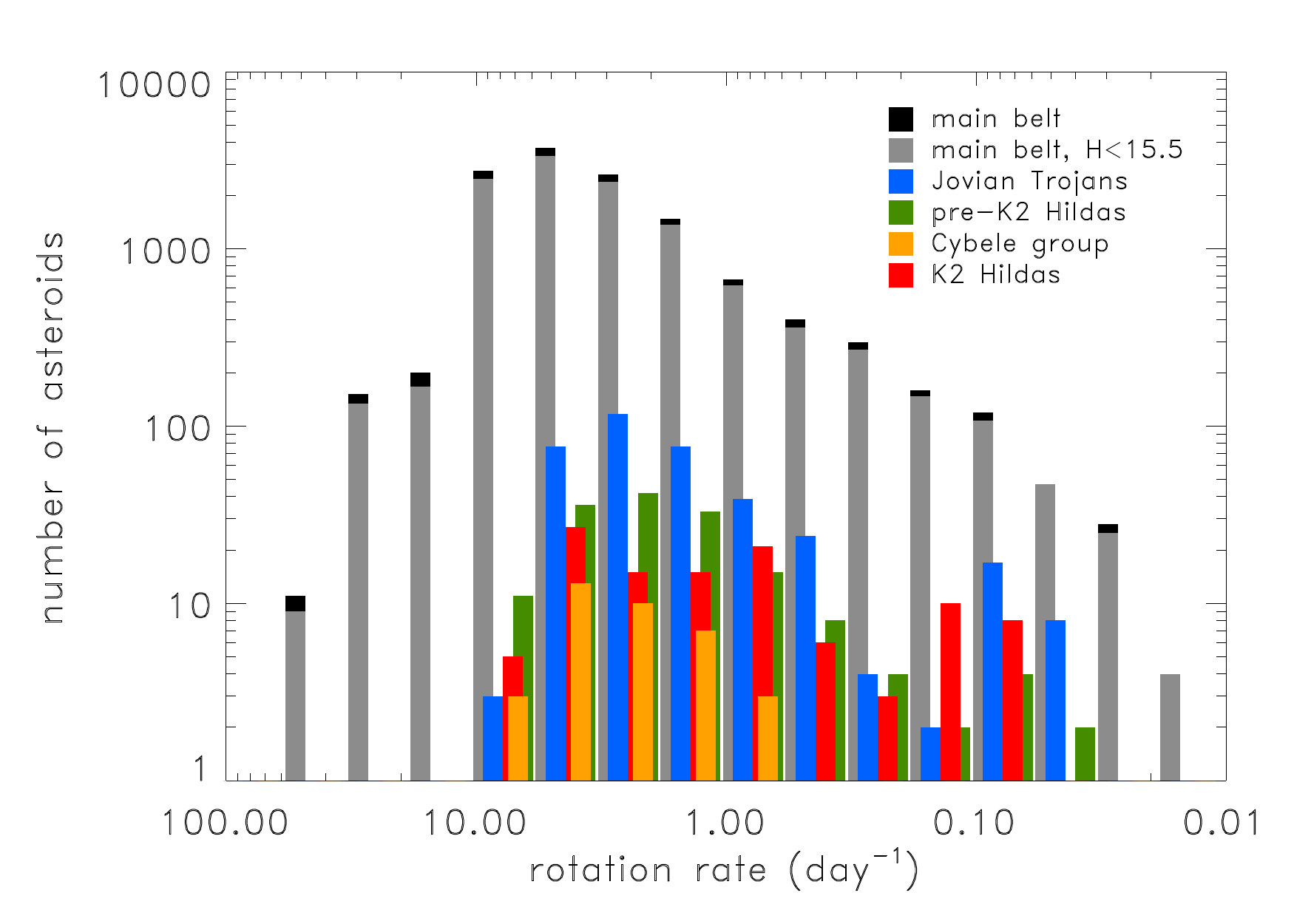}
\caption{Top: Rotation rate -- absolute magnitude distribution of K2 Hilda asteroids (large red dots) compared with Earth-based Hilda observations prior the K2 sample (green dots), Jovian Trojan (blue points), Cybele asteroids (orange dots) and the main belt (gray points). Bottom: The histogram of rotation rates, with the same color coding as in the above panel, but with black and gray bars marking the full main belt sample and the asteroids with H\,$\leq$\,15.5\,mag, respectively. In this plot the sample of Cybele asteroids is restricted to H\,$\leq$\,15.5\,mag, too.}
\label{fig:per-absbrightness}
\end{figure}

\begin{figure}
\includegraphics[width=\columnwidth]{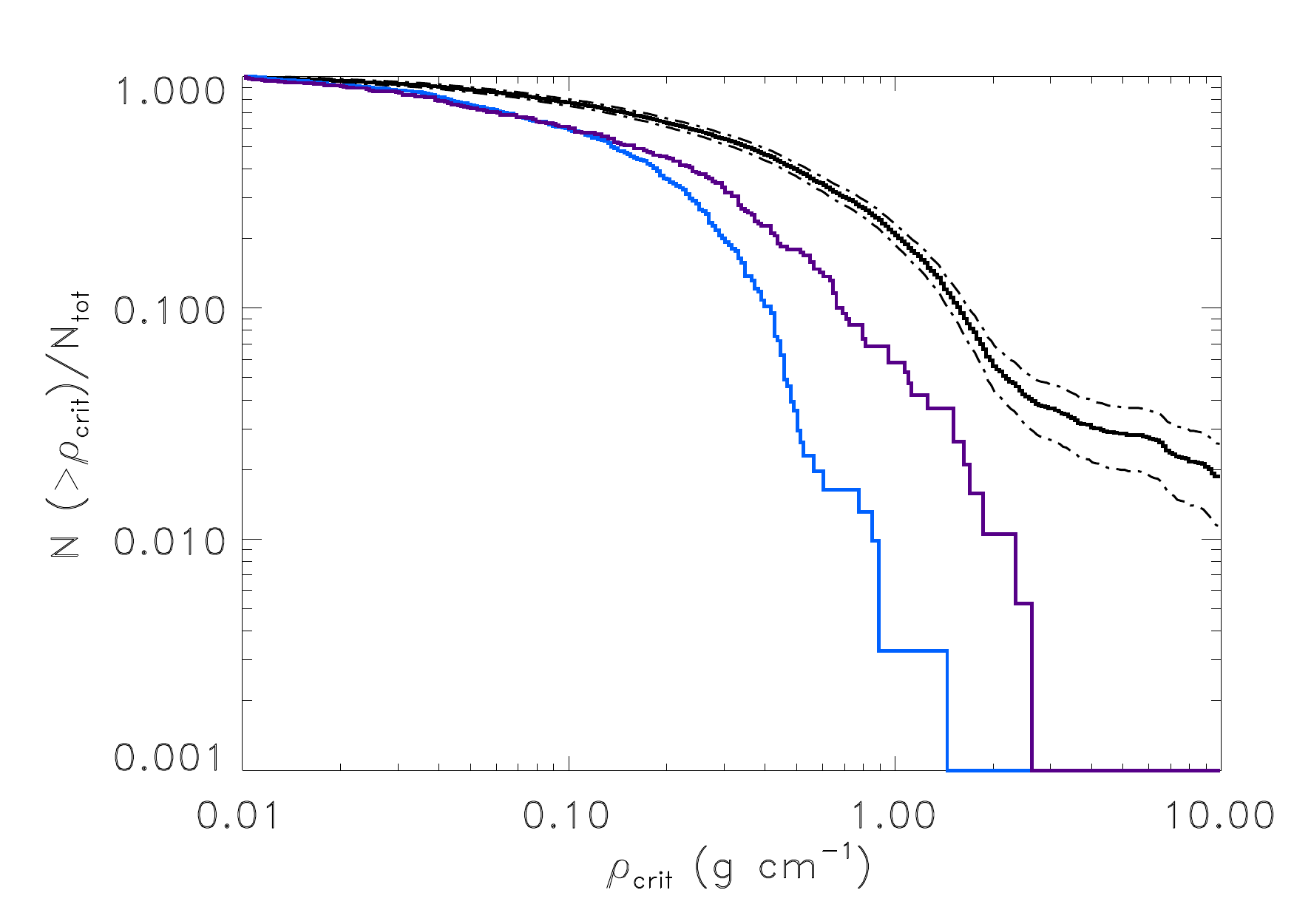}
\caption{Cumulative number of asteroids with critical 
densities ($\rho_{crit}$) above a specific value for main belt (black curve, restricted to H$_V$\,\,15.5\,mag), 
Jovian Trojans (blue) and Hilda asteroids (purple, including both pre-K2 and K2 data). Dash-dotted curves around the 
main belt (black) curve represent the 1-$\sigma$ confidence intervals.}
\label{fig:cumulative-density}
\end{figure}

\section{Analysis}

In the upper panel of Fig.~\ref{fig:per-absbrightness}, we plot the rotation rates against the absolute brightness of the observed K2 Hilda asteroids, in comparison with the Main Belt and Trojan asteroids and previous Hilda observations. A similarity is suggested between Hilda and Trojan asteroids, most importantly in the range of very slow rotation. The $P>5$~d (rotation rate $<$0.2~d$^{-1}$) wing is populated by many members of both the Hilda and the Jovian Trojan families, and almost completely avoided by Main Belt asteroids. Even, the two samples overlap in the $12<H_V<13.5$ where a direct comparison is possible. The K2 sample represents a large number of fainter asteroids, which can appear at higher rotation rates than the larger asteroids, which reflects the size dependence of the break-up velocity. In this sense, Hildas and Jovian Trojans also follow a similar distribution, suggesting a higher porosity of Hilda asteroids. This is again a similarity between Trojan and Hilda asteroids. 


In the lower panel of Fig.~\ref{fig:per-absbrightness} we compare the spin frequency distribution of Hilda asteroid with that of minor bodies from the main belt and Jovian Trojans. Data for these latter populations, as well as for the pre-K2 Hilda measurements are taken from the Light Curve Database \citep[LCDB,][]{Warner2009}. Cybele asteroids were selected by their osculating orbital elemelents, taken from the Minor Planet Center MPCORB.DAT file, and cross-matched with the Light Curve Database. We required 3.28\,au\,$\leq$\,$a$\,$\leq$\,3.70\,au, $e$\,$\leq$\,0.3 and $i$\,$\leq$\,25\degr. Only 39 Cybele asteroids were identified with known rotation periods and absolute brightness below 15.5\,mag, making it difficult to draw conclusions from the statistics in this group. For the main belt sample we excluded the data of asteroids families and included minor bodies that were specifically assigned to the main belt (family codes of MB-I, MB-M and MB-O).

The spin frequency distribution of Hildas and Jovian Trojans are similar, with a prominent secondary peak at f\,$\sim$\,0.1\,d$^{-1}$, not seen in the main belt population. If a double-Maxwellian distribution is fitted to the full Hilda sample the two peaks of the distribution are at f\,=\,2.5\,d$^{-1}$ (P\,=\,9.6\,h) and f\,=\,2.5\,d$^{-1}$ (P\,=\,170\,h). (Here it has to be noted that the Maxwellian shape is plausible if the group is collosionally relaxed. We do not know this, neither can we test the Maxwellian character of the distribution due to the small sample size. We just consider the Maxwellian fits because they nicely fit the distribution, give a firm statistics for a characteristic rotation period, but do not interpret this fit in the context of dynamical (non) relaxed state.) 

Using the definition of slow and fast rotations by \citet{Pravec2000} the ratio of slow rotators (f\,$\leq$\,0.8\,d$^{-1}$) is $>$20\,\% both among Hildas and Jovian Trojans, notably larger than in the main belt (9.5\%, see also Table~\ref{table:fstat}). Similarly, very slow (P\,$\geq$\,100\,h) rotators are also notably abundant in these two resonant populations. A direct comparison is possible in the size range ($H_{\rm V}$=10--14) where all three samples overlap. A factor of 2-3 larger number of slow rotators among K2 Hildas compared with the mostly ground-based Jovian Trojan and pre-K2 Hilda periods show that space surveys with long, uninterrupted time series photometry may more easily detect long period light curves. This has already been the case for the sample of main belt asteroids observed with K2 \citep{2018ApJS..234...37M} -- the existence of a larger number of slow rotators in the main belt is expected to be confirmed by the TESS Space Telescope due to the longer coverage of uninterrupted observations, and also because the one-day alias periods do not emerge in uninterrupted data, and the period determination will be unique even for very slow rotators \citep{2018PASP..130k4503P}. Therefore, the comparison of the fraction of slow rotators among Trojans, Hildas and the Main Belt must be revisited after the completion of the TESS survey.

Fast rotators (f\,$\geq$\,7\,d$^{-1}$) are notably more common in the main belt (13.3\%) than among Hildas and Jovian Trojans (2.1 and 0.2\%, respectively). In fact, fast rotators are almost completely missing from the Jovian Trojan population. It has to be noted that the Hilda, Trojan and MB samples cover different size ranges in Fig. 3: Trojans down to $H_{\rm V}$=13.5, Hildas down to $H_{ \rm V}$=15.5 and MB asteroids down to $H_{\rm V}$=19, and the histogram in Fig 3 lower panel mixes together asteroid samples extending to differing small size limits. The upper panel of Fig. 3 may also suggest that the distribution of fast rotating MB asteroids really extends to higher rotation rates than Hildas and Trojans even at the area where all samples overlap, in the $H_{\rm V}$=13--14.5 range. This was previously explained by the likely low bulk density of these objects \citep{2017A&A...599A..44S}. Using the rotation periods and light curve amplitudes we calculated the distribution of critical densities in the main belt, Jovian Trojan and Hilda populations (Fig.~\ref{fig:cumulative-density}). 

To account for the much larger number of main belt asteroids than Hildas and Jovian Trojans in our sample we randomly selected the same number of asteroids from the main belt sample as in the Jovian Trojan sample multiple times, and calculated the mean curve and the $\pm$1$\sigma$ confidence intervals (solid and dash-dotted curve in Fig.~\ref{fig:cumulative-density}).
The curves indicate a fast extinction of Jovian Trojans -- they can be rarely found at $\rho_{crit}$\,$>$\,0.5\,g\,cm$^{-3}$. Hildas, on the other hand, can easily reach  $\rho_{crit}$\,$\approx$\,2\,g\,cm$^{-3}$, similar to the breakup limit of main belt asteroids. The high density wing of the main belt is not seen in the two resonant populations. The existence of high critical density Hildas has an important implication. While the low critical densities of Jovian Trojans is in agreement with their outer solar system origin \citep[as discussed e.g.][]{2017A&A...599A..44S}, the $\sim$\,2\,g\,cm$^{-3}$ critical density of some Hildas indicate that their building material should be closer to that of main belt asteroids. This suggests a mixed, partly main belt, partly outer solar system, origin of Hildas. 

\begin{table*}
\centering
\begin{tabular}{|ccccccc|| ccc |}
\hline
     &  MB   & MB$_{15}$ & JT & Cy. & pre-K2 Hil. & all Hil. & K2 MB & K2 JT & K2 Hil.   \\
\hline
        N  & 13072  & 7874      & 401   & 39 & 187      & 298    &81& 56 & 111        \\
f$_m$ [d$^{-1}$]& 3.53 & 3.95      & 2.12  & 2.95   & 2.63     & 2.09   &3.63 & 1.76 & 1.26       \\  
P$_m$ [h]    & 6.79   & 6.07      & 11.3  & 8.11 & 9.13     & 11.5  &6.60 & 13.58& 19.02     \\
\hline
N$_{f}$    & 1930     & 1057      & 1     & 0 & 7          & 9    & 11 & 2 & 2        \\
r$_{f}$ [\%] & 14.8   & 13.4      & 0.3   & 0 & 3.7       & 3.0   & 13.6 & 3.7 & 1.8       \\
\hline
N$_{s}$   & 1445      & 908      & 79     &  2       & 29         & 71   & 5 & 14 & 42       \\ 
r$_{s}$ [\%]  & 11.1   & 11.5       & 19.7   &  5.1       & 15.5      & 23.8  & 6.2 & 25.9 & 37.8      \\
\hline
N$_{vs}$   & 488     & 320       & 29    & 0 & 11        & 31    & 0 & 9 & 20        \\ 
r$_{vs}$ [\%] & 3.7  & 4.1       & 7.2   & 0 & 5.9      & 10.4  & 0 & 16.7 & 18.0      \\
\hline
\end{tabular}
\caption{Summary table of median rotation rates 
(f$_m$, and the corresponding period P$_m$), and the number of slow and fast rotating asteroids in the main belt (MB), 
Jovian Trojan (JT), Cybele (Cy) and Hilda populations. We defined fast rotators (subscript 'f') 
as f\,$\geq$\,7\,d$^{-1}$ (P\,$\leq$\,3.43\,h), slow rotators ('s') as f\,$\leq$\,0.8\,d$^{-1}$ (P\,$\geq$\,30\,h) 
and very slow rotators ('vs') as f\,$\leq$\,0.24\,d$^{-1}$ (P\,$\geq$\,100\,h). The Cybele and Main Belt populations are restricted to H\,$\leq$\,15.5\,mag.}
%
%
%
%
\label{table:fstat}
\end{table*}


Among Hildas with the longest periods, the typical amplitude is in the 0.4--1.0 magnitude range. As suggested in the case of 
slowly rotating, high amplitude Trojans (\cite{2017A&A...599A..44S} and references therein), the high-amplitude, long-period Hilda 
asteroids can be considered as potential binaries. 

Slow rotation of asteroids, however, could also be caused by other effects. This could be, e.g. the disintegration of high mass ratio ($\sim$1:5) binaries through rapid transfer of rotational energy of the primary into the orbit of the secondary due to the irregular shape and gravity field of the primary \citep{Harris2002}. Small (D\,$\lesssim$\,20\,km) main belt asteroids show a significant deviation from the Maxwellian distribution seen among larger asteroids. This can be well explained by a relaxed YORP evolution  \citep{Pravec2008,Vok2015} for spin rates of f\,$\approx$\,1--10\,d$^{-1}$. In this sense Hilda asteroids may also be susceptible to the YORP effect; Hildas in the K2 sample have H$_V$\,$\approx$\,10.5--15.5\,mag or D\,$\approx$\,3--40\,km. However, in the whole Hilda sample, there are asteroids with slow rotation periods with D\,$>$\,40\,km (H\,$<$\,10\,mag) and the rotation periods in Cluster 3 are notably longer than that explained by the model of \citet{Pravec2008}. 



\begin{table}
    \centering 
    \footnotesize
    \begin{tabular}{lr@{$\pm$}lr@{$\pm$}lcr@{.}lr@{.}l}
\hline
    Number & \multicolumn{2}{c}{$r-i$} & \multicolumn{2}{c}{$i-z$}& Group & \multicolumn{2}{c}{P[h]} & \multicolumn{2}{c}{Amp} \cr 
    \hline 
1748 &0.21  &  0.11  &  0.14  &  0.05  &  R          & 6&005 & 0&09\cr 
3655$_{\rm A}$ &0.20  &  0.03  &  0.20  &  0.03  &  R & 77&047 & 0&19 \cr
7394 &0.24  &  0.02  &  0.15  &  0.02  &  R         & 4&836 & 0&16 \cr
8550 &0.11  &  0.04  &  0.16  &  0.07  &  LR        & 37&209 & 0&28 \cr
15278 &0.28  &  0.02  &  0.14  &  0.03  &  R        & 39&160 & 0&26 \cr
15626 &0.16  &  0.01  &  0.10  &  0.02  &  LR       & 13&544 & 0&75 \cr
16232 &0.13  &  0.04  &  0.03  &  0.05  &  LR       & 152&4 & 0&69 \cr
16843 &0.25  &  0.02  &  0.18  &  0.04  &  R        & 310&0 & 0&28 \cr
20628$_{\rm B}$ &0.23  &  0.02  &  0.18  & 0.03  &  R & 320&0 & 0&95 \cr
27561 &0.20  &  0.02  &  0.18  &  0.03  &  R        & \multicolumn{2}{c}{---} & $<$0&1 \cr
43818 &0.21  &  0.03  &  0.21  &  0.05  &  R        & 21&64 & 0&29 \cr
46302 &0.12  &  0.04  &  0.08  &  0.05  &  LR       & 7&479 & 0&14 \cr
51930 &0.22  &  0.04  &  0.19  &  0.02  &  R        & 52&288 & 0&37 \cr
57027$_{\rm 1}$ &0.27  &  0.03  &  0.14  &  0.06 &  R & 4&800 & 0&15 \cr
57027$_{\rm 2}$ &0.27  &  0.03  &  0.14  &  0.06 &  R & 25&798 & 0&12 \cr
60381$_{\rm A}$ &0.21  &  0.02  &  0.13  &  0.02 &  R & 5&897 & 0&17 \cr 
78159 &0.25  &  0.03  &  0.18  &  0.06  &  R        &15&404 & 0&22 \cr
90704$_{\rm 1}$ &0.27  &  0.04  &  0.13  &  0.09 & R & 4&760 &0&15 \cr
90704$_{\rm 2}$ &0.27  &  0.04  &  0.13  &  0.09  &  R &23&762 &0&27 \cr
92281 &0.15  &  0.06  &  0.10  &  0.11  &  LR        & 65&529 & 0&09 \cr
98002 &0.22  &  0.05  &  0.08  &  0.1  &  R          & 7&557 & 0&25 \cr
99276 &0.12  &  0.07  &  0.06  &  0.16  &  LR        & 5&239 & 0&07 \cr
111995 & 0.24  &  0.04  &  0.11  &  0.07  &  R       & \multicolumn{2}{c}{---} & $<$0&1 \cr
117106 &0.11  &  0.05  &  0.12  &  0.11  &  LR       & 32&090 & 0&21 \cr
117113 &0.15  &  0.03  &  0.11  &  0.04  &  LR       & 136&908 & 0&40\cr
119918 &0.12  &  0.02  &  $-$0.01  &  0.05  &  LR    & 5&006 & 0&45 \cr
161606 &0.30  &  0.06  &  0.15  &  0.13  &  R        & 42&590 & 0&21\cr
174077$_{\rm 1}$ &0.33  &  0.06  &  0.04  &  0.08  & R & 3&406 & 0&13\cr 
174077$_{\rm 2}$ &0.33  &  0.06  &  0.04  &  0.08  & R & 24&328 & 0&17\cr
174089 &0.27  &  0.03  &  0.10  &  0.05  &  R        & 148&194 & 0&86\cr
177943 &0.12  &  0.05  &  0.08  &  0.1  &  LR        & 19&020 & 0&27 \cr
193449 &0.25  &  0.06  &  0.16  &  0.33  &  R        & 21&111 & 0&20\cr
202992 &0.28  &  0.06  &  0.05  &  0.11  &  R        & \multicolumn{2}{c}{---} & $<$0&1 \cr
207638 &0.29  &  0.06  &  0.07  &  0.13  &  R        & 37&354 & 0&28 \cr
207644 &0.18  &  0.05  &  $-$0.04  &  0.11  &  LR    & 2&689 & 0&23\cr
217032$_{\rm C}$ &0.25  &  0.09  &  0.22  &  0.16  & R & 7&704 & 0&42 \cr
230294 & 0.08  &  0.04  &  0.08  &  0.07  &  LR      & 11&440 & 0&57 \cr
236209$_{\rm 1}$ &0.06  &  0.02  &  0.07  &  0.04&  LR & 6&095 & 0&15 \cr
236209$_{\rm 2}$ &0.06  &  0.02  &  0.07  &  0.04&  LR & 18&480 & 0&15 \cr
341841 &0.16  &  0.04  &  $-$0.06  &  0.08  &  LR    & 4&030 & 0&18\cr
\hline
    \end{tabular}
    \caption{SDSS colors, Red (R) / Less Red (LR) group identifications and period, amplitude rotation parameters for K2 asteroids with SDSS counterparts. The remarks in the subscripts are as follows. 
    {\bf A}: Three humps in the light curve;
    {\bf B}: Variable light curve shape, possible tumbler; 
    {\bf C}: Period and amplitude from Waszczak et al. (2015);
    {\bf 1}: First period of a tumbler;
    {\bf 2}: Second period of a tumbler.
    }
    \label{tab:k2sdss}
\end{table}

Hilda asteroids of different taxonomy classes are believed to be of different origin --  C and X types are likely related to the main belt while P and D types to the Jovian Trojan swarms -- and then evolved as a member of the Hilda group \citep[see e.g.][]{2017A&A...599A..44S}. 
A major question related to the present survey was to decide whether the Hildas from different origin preserved the initial rotational properties, or they evolved toward a common distribution, characteristic of Hilda asteroids. Since there are few Hildas with identified R and LR taxonomy in our sample, the direct comparison of the distributions are inconclusive. Instead, we compared the R and LR subset of Hildas to the Trojan and Main Belt samples in our previous K2 publications \citep{2017A&A...599A..44S, 2016A&A...596A..40S}. 

In Table \ref{tab:k2sdss}, we present the derived periods and amplitudes for those Hildas that had SDSS detections, too (either from MOC, or from our search from the PhotoObj files). Here we give the $r-i$ and $i-z$ color indices, the R/LR classification of the Hildas, and the period and amplitude as involved in the following analysis. (In case of multiple observations, the periods derived in the different campaigns were averaged, and the largest observed amplitude was considered. 
In case of tumbler asteroids with multiple periods, the shorter period was considered. In the case of 185290, the dominant period in C08 and C13 agree within 2\%{}, 
although visually C08 is better with three humps, and C13 is better looking with two humps. Since this issue cannot be decided from the current data, we included the two 
hump solution in the analysis. 

\begin{figure}
\includegraphics[viewport=2 15 406 341,width=\columnwidth]{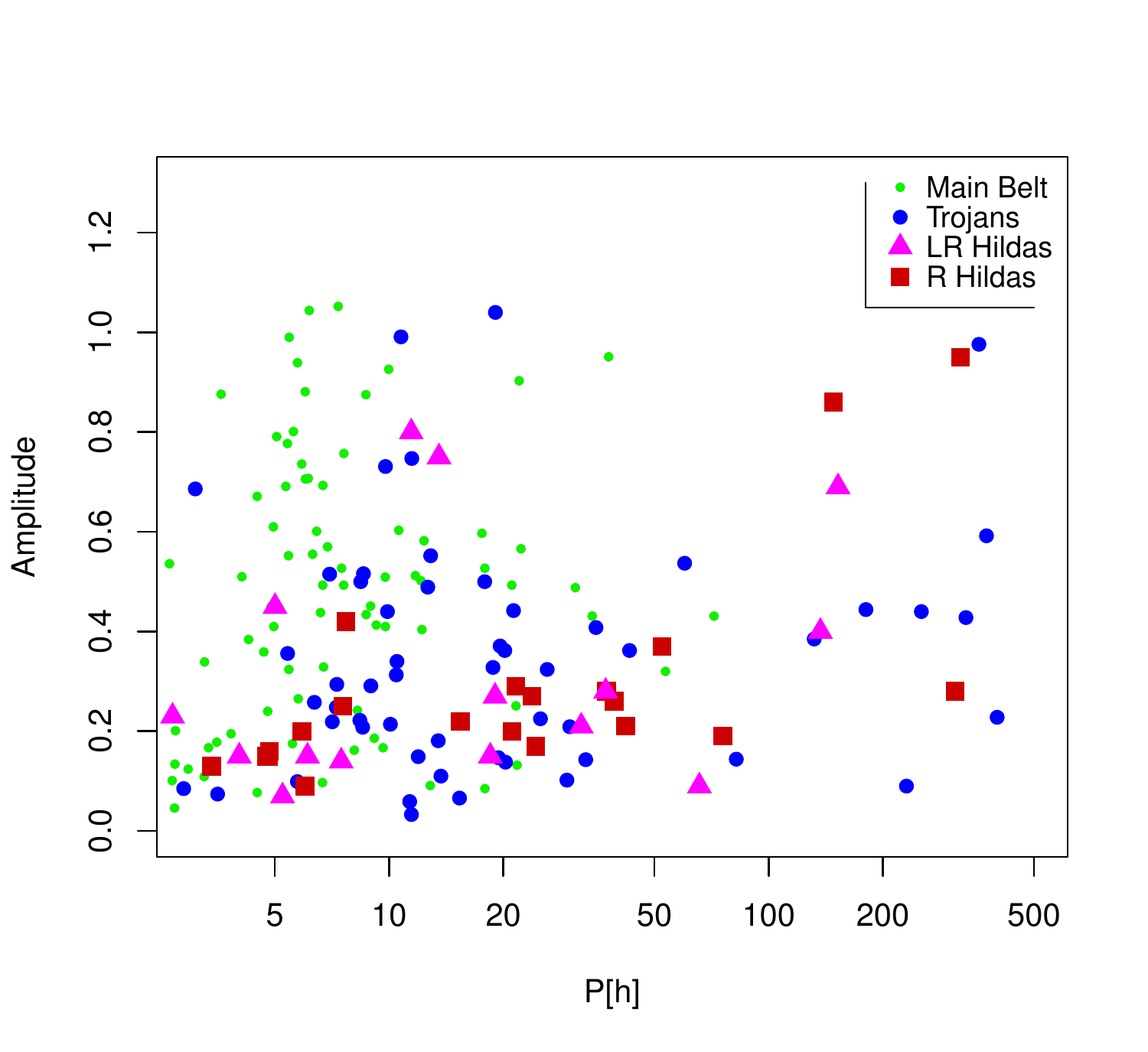}
\caption{The distribution of LR and R groups of K2 Hilda asteroids, compared to the {K2} sample of Main Belt and Trojan asteroids.}
\label{fig:populations}
\end{figure}

In Fig.~\ref{fig:populations}, we plot the LR and R Hildas in the period--amplitude space, in comparison with main belt and Jovian Trojan asteroids in the K2 data. The distribution of R and LR Hildas and Jovian Trojan asteroids are apparently very similar, and we tested this similarity with the energy distance test as described by \cite{szekelyrizzo2004}.
This test is a powerful alternative of the Kolmogorov-Smirnov test in arbitrary multiple dimensions, but since it avoids dimension reductions, it preserves the full information. To measure the distance, the data set has to be normalized along the independent dimensions. Hence, we normalized the amplitude in all distributions with the standard deviation of amplitude distribution of the (unified) Hilda asteroids, and the similar recipe for the period was followed, too. The test is evaluated using a so-called jacknife approach, which differs from methods involving the complete bootstrap. Here, only a single pair of random elements from both original samples are chosen and their assignment is commuted. This way, all test samples differ from the original distribution by a split of one single element only. If the samples are different, this change usually decreases the energy distance, because information is lost when distributions of different pattern are mixed together. The test statistics $\alpha$ will be the fraction when changing the single elements led to an increase of the energy distance, therefore significantly differing distributions are characterized by little values of $\alpha$. \cite{szekelyrizzo2004} suggests that distributions are significantly different when $\alpha<0.1$.

We found that the rotational properties of R type Hilda asteroids differ very significantly from that of Main Belt asteroids ($\alpha$=0.001). Moreover, the rotation properties of Hilda asteroids are seemingly {\it farther} from the main belt than that of the Trojans from the main belt (this latter test for different background distribution had a $\alpha$=0.01 value). The period-amplitude distribution of LR type Hildas is likely distinguishable from Main Belt asteroids ($\alpha$=0.05), but indistinguishable from Trojans. Also, the unified sample of all Hildas is indistinguishable from the Trojans by rotation properties in a direct comparison.

This comparison results in a somewhat surprising picture. The R group of Hilda asteroids (D and P type for taxonomy) follows a similar period and amplitude distribution as the Jovian Trojans, which is an indication of their origin from the Trojan swarms, and that later they preserved the initial rotation properties. On the other hand, the period and amplitude distributions of the LR group (mostly C and X type) Hilda asteroids {\it also} resemble that of the Trojan asteroids, and not the main belt. This can either mean that (1) there is a recognisable distribution of rotation periods and amplitudes in the Hilda group, and all asteroids evolved to this distribution by now, regardless of their origin; and this distribution is close to that of the Jovian Trojans; or (2) all types of Hilda asteroids could preserve their original rotation properties, but the LR type Hildas originated from a part of the outer Main Belt where the rotation amplitudes and periods were unlike in the major part of the Main Belt, but very close to the Jovian Trojans instead; (3) the K2 observations of Main Belt asteroids covered a shorter average time span and is truncated for very slow rotation periods, preventing us from directly comparing the K2 Hildas and Trojan asteroids. The answer has to be postponed to later surveys such as the Solar System objects in TESS data, or the LSST. 



\subsection{Hilda asteroids with double periods}


In the K2 sample we found an unusually high ratio of asteroids with double periods. Similar observations of double periods are usually interpreted as an outcome of a double asteroid, or a single asteroid in the state of tumbling \citep{2002aste.book..517P}. 

Asteroid tumbling is explained by the misalignment of the axis of instantaneous rotation and the angular momentum vector. Therefore, the axis of the instantaneous rotation traces the herpolhodie cone (with the axis of symmetry being the angular momentum vector), and the axis of maximal inertia ($c$ axis in the triaxial shape approximation of the body shape) traces along the nutation cone (whose axis of symmetry is also the angular momentum vector, but the cone angle can be different). Thus, the body can be interpreted as rotating around two axes, and two periods are observed in the light curves. 

Asteroid tumbling is damped by tidal forces from distant bodies and internal stress waves \citep{2002aste.book..517P}. The tumbling state of a free Solar System body ceases in the time scale of $10^5$~yr, and is considered to be a transient state of rotation. Tumbling can be straightforwardly triggered by an energetic impact between asteroids. It has to be noted that the YORP effect can also excite tumbling, and since YORP can act continuously, this kind of tumbling does not decay (decays or excites slowly together with the YORP induced evolution of the rotation axis \citep{2009Icar..200..514S}).

Another usual reason of a second period is a large companion around the asteroid. We can observe the rotation of the two bodies as two periods in the light curve if the companion is not much smaller than the primary body. 

In the case of the tumbling, the period ratio depends on the moments of inertia, the period ratio equals $\sqrt{(\Omega_3/\Omega_1 -1)(\Omega_2/\Omega_1 -1)}$. From the large size and the low observed amplitude (0.12--0.17) of the Hilda asteroids with double periods, likely $\Omega_{\rm 2,3}$ are close to $\Omega_{\rm 1}$, and the ratio of the rotation and precession periods are predicted to differ significanlty. Indeed, the observed period ratios are roughly 3, 5 and 7.2 for asteroids 236209, 90704 and 174077, respectively, which is compatible with tumbling.

The binarity scenario cannot be excluded either, because the fast signal in the light curves can be considered as the rotation of the main body, while the period of the slow signals are compatible with the bound revolution of a companion. Most importantly, in the case of 236209 and 90704, where the period ratios are very close to integers 3 and 5, respectively, pointing towards a possible  spin-orbit resonance. 

In the case of a very long period asteroid (20628) 1999~TS40, P$\approx$320 h, only two rotation cycles were observed, but the notable light curve variations are characteristic of a tumbler asteroid. The tumbling nature of (20628) 1999~TS40 was first proposed by \citet{Pravec2005}, later debated by \citet{Pravec2014} but confirmed by \citep{2018MPBu...45..262W}. The long period continuous light curve of { K2} proves the long time systematic light curve variations, as an evidence for tumbling.

All tumblers have been identified in SDSS (90704 1988~RO12, 174077 2002~FP21, and 236209 2005~WT122, besides of the above discussed 20628 1999~TS40). Of the tumblers, (236209) 2005~WT122 is the only LR-type Hilda asteroid, while the three other tumblers belong to the R group. 

We identified one asteroid, (57027) 2000 UB59, that is most likely an asynchronous binary, i.e., a larger primary with a small companion, and the rotation of the larger component has not synchronised with the orbital period. The light curve of 57025 shows both a fast, rotation-like signal ($P_{rot} = 4.80$ h), and a longer variation characterized by short, distinct eclipses, with a maximal depth of 0.12 mag ($P_{orb} = 25.80$ h). 

\subsection{Hilda asteroids with triple light curve symmetry}

In the { K2} Hilda sample, we found 4 asteroids exhibiting a trimodal light curve with three humps (three minima and three maxima) during one rotation: 1748, 3655, 60381, 249182. (We note that C08 observations of 185290 can also be better interpreted as three humps with 85~h rotation period, while C13 observations are more symmetric, can be compatible with either 57.5 or 85.3~h period.)  Majority of light curves with two minima/maxima in one cycle can be well fitted with a rotating ellipsoid, but these trimodal light curves represent a prominent deviation from the usual case. The simplest shape that can be seen to generate a trimodal light curve is a tetrahedron with an inclined aspect of the rotation angle and/or phase angle of incident light. Such shape is not necessarily built up by mergers, but a merger outcome can be similar. Therefore, the large fraction of light curves with three humps can also be a sign of a significant fraction of binary asteroids and merged asteroids in the Hilda group.

\subsection{Hilda asteroids with very long periods}

As presented in Fig~\ref{fig:per-absbrightness} and Table~\ref{table:fstat} the Hilda period distribution has a very heavy wing towards long periods. Indeed, the distribution (Fig.~\ref{fig:per-absbrightness}) indicates that a notable characteristic of the rotation properties is the surprisingly large fraction of long-period Hilda asteroids. About 38\%{} of K2 Hildas have $P_{\rm rot}$\,$\geq$\,30\,h, and 18\%{} of Hildas reside in the $P_{\rm rot}>$100~h regime. A similar overpopulation of slow rotators was observed in the Jovian Trojan K2 sample. On the contrary, none of the 120 Main Belt K2 asteroids showed a rotation period above 100~h (see also Table \ref{table:fstat}.).


In \cite{2017A&A...599A..44S} we concluded that the Trojan swarms contain a considerable fraction of unusual asteroids -- slow rotators, tumblers and trimodal light curves -- implying a large number of binaries in the Trojan swarms.  Although the rotation periods and amplitudes are very similar in the case of Hilda and Trojan asteroids, it is interesting that the ratio of unusual asteroids is even larger in the Hilda group than in the Trojan swarms, the most significant difference is the rich population of long period asteroids especially in the case of very slow rotators ($P>100$~h) in the Hilda group which is true in total numbers, Fig. 3. lower panel, and also true in the overlapping size region, Fig. 3 upper panel. These findings suggest a general similarity of Trojan and Hilda asteroids for dynamical properties, but, interestingly, the binary fraction in the Hilda group is likely even larger than in the Trojan swarms. The same was concluded by \cite{2015ApJ...799..191S} from NEOWISE data, which we confirm from K2 observations. At least from this aspect, the Hilda group is not an intermediate mix between the Trojans and the Main Belt, but it represents the end of the range.

\section{Summary}
In this paper we analysed 125 individual light curves of Hilda asteroids from the {K2} mission, and concluded the following results.
\begin{itemize}
  \item{} The period-amplitude distribution of Hilda asteroids is very similar to the Trojan populations. In fact, rotation properties of Hildas and Trojans are statistically indiscernible.
  \item{} The rotation parameters of Hildas are distinctly different from that of the Main Belt, as there are only a low number of fast rotator Hildas, however, this ratio in even lower for the Jovian Trojans. This is also true for LR type Hildas, although they are expected to originate from the Main Belt.
   \item{} Both Hilda subgroups with Trojan-like (R group, D and P) and Main Belt-like (LR group, C and X) taxonomy share very similar rotation properties.
   \item{} The median rotation period of K2 Hilda asteroids is 20.7 hours, the mean period is 54.8 hours. Both values prominently exceed that of the known Main Belt asteroids. There is an unprecedentedly large fraction ($\approx$ 18\%) of extreme slow rotators ($P$\,$>$\,100\,h).
   \item{} We found 4 asteroids with double periods, where binarity is the most reasonable explanation in 2 cases. The other 2 cases can be a result of either binarity or tumbling.
   \item{} We identified 5 asteroids with three maxima in the light curve.
   \item{} We found that the binarity rate of Hildas is among the highest in the asteroid belt, likely exceeding that of the Trojan swarms, too.
\end{itemize}

\acknowledgments
This paper includes data collected by the K2 mission. Funding for the K2 mission is provided by the NASA Science Mission directorate. This  project  has  been  supported  by  the  Hungarian National Research, Development and Innovation Office (NKFIH) grants K-119517, K-115709, K-125015, GINOP-2.3.2-15-2016-00003, the Lend\"ulet Program of the Hungarian Academy of Sciences, project No. LP2018-7/2019, and the City of Szombathely under Agreement No.\ 67.177-21/2016. L.M. was supported by the Premium Postdoctoral Research Program of the Hungarian Academy of Sciences. The data presented in this paper were obtained from the Mikulski Archive for Space Telescopes (MAST). STScI is operated by the Association of Universities for Research in Astronomy, Inc., under NASA contract NAS5-26555. Support for MAST for non-HST data is provided by the NASA Office of Space Science via grant NNX09AF08G and by other grants and contracts. The authors thank the hospitality the Veszpr\'em Regional Centre of the Hungarian Academy of Sciences (MTA VEAB) where parts of this project were carried out.

%

\vspace{5mm}
\facilities{Kepler}
\software{FITSH \citep{fitsh}, gnuplot}


\begin{appendix}


\startlongtable
\begin{deluxetable*}{lcccrrrrrr}
\tabletypesize{\tiny}
\tablecaption{Hilda asteroids observed by K2, ordered by the campaign of the observation. The columns are: (1) the number and name/designation; (2) the reference to the K2 campaign when the asteroid was observed; (3-4) start and end date in Julian Date; (5) the length of obervations in days, (6) the number of frames where the asteroid was detected; (7) the duty cycle of the observations (ratio of useful cadences and all cadences over the time span of the observations); (8) the $r$ heliocentric distance in AU; (9) $\Delta$ K2--object distance in AU; (10) the phase angle of the observation from K2. } 
\tablehead{\colhead{Name}  &  \colhead{Cam.}  &  \colhead{Start}  &  \colhead{End}  &  \colhead{Length}  &  \colhead{\#frame}  &  \colhead{Duty}  &  \colhead{$r$} & \colhead{$\Delta$} & \colhead{$\alpha$} 
}
\colnumbers
\startdata
(113224) 2002 RN121 & C06 & 2457220.6050 & 2457230.9648 & 10.360 & 481 & 0.946 & 4.242...4.255 & 3.469...3.570 & 10.342...11.887 \\
(15626) 2000 HR50 & C06 & 2457262.4121 & 2457270.3199 & 7.908 & 337 & 0.868 & 4.308...4.312 & 4.163...4.280 & 13.962...14.061 \\
(3655)  Eupraksia & C06 & 2457233.2534 & 2457253.1966 & 19.943 & 663 & 0.678 & 4.729...4.741 & 4.077...4.330 & 10.600...12.303 \\
(39301) 2001 OB100 & C06 & 2457261.1452 & 2457271.5459 & 10.401 & 434 & 0.850 & 4.397...4.413 & 4.128...4.263 & 13.555...13.738 \\
(46302) 2001 OG13 & C06 & 2457282.1510 & 2457293.0217 & 10.871 & 453 & 0.849 & 3.678...3.700 & 3.881...3.999 & 14.756...15.543 \\
(51930) 2001 QW127 & C06 & 2457217.4582 & 2457228.3289 & 10.871 & 506 & 0.949 & 4.204...4.221 & 3.426...3.530 & 10.288...11.977 \\
(78159) 2002 NA28 & C06 & 2457223.4861 & 2457229.6571 & 6.171 & 274 & 0.904 & 4.487...4.491 & 3.741...3.806 & 10.256...11.085 \\
(8550)  Hesiodos & C06 & 2457280.5163 & 2457288.3219 & 7.806 & 355 & 0.926 & 4.696...4.706 & 4.895...4.995 & 11.783...12.247 \\
(99276) 2001 QC20 & C06 & 2457226.9598 & 2457247.0052 & 20.045 & 925 & 0.941 & 3.926...3.964 & 3.239...3.452 & 12.153...14.605 \\
(185290) 2006 UB219 & C08 & 2457405.2838 & 2457434.3608 & 29.077 & 1256 & 0.881 & 4.255...4.288 & 3.585...3.960 & 10.010...9.969 \\
(202992) 1999 VG135 & C08 & 2457392.0633 & 2457412.4356 & 20.372 & 371 & 0.371 & 2.751...2.760 & 1.955...2.164 & 13.912...18.436 \\
(203157) 2000 WC140 & C08 & 2457392.0633 & 2457418.3818 & 26.319 & 1128 & 0.875 & 3.194...3.241 & 2.448...2.721 & 11.805...16.667 \\
(20628) 1999 TS40 & C08 & 2457394.4949 & 2457426.5143 & 32.019 & 1475 & 0.940 & 3.171...3.204 & 2.377...2.720 & 10.995...16.955 \\
(20630) 1999 TJ90 & C08 & 2457392.0633 & 2457416.5019 & 24.439 & 1188 & 0.992 & 2.884...2.900 & 2.083...2.338 & 12.752...17.936 \\
(402869) 2007 RY194 & C08 & 2457393.8001 & 2457427.1069 & 33.307 & 1595 & 0.977 & 3.436...3.445 & 2.632...3.035 & 10.482...15.849 \\
(148227) 2000 DP99 & C10 & 2457582.5862 & 2457589.3701 & 6.784 & 317 & 0.951 & 3.497...3.505 & 2.781...2.864 & 14.038...15.033 \\
(152133) 2004 TN126 & C10 & 2457582.5862 & 2457612.9914 & 30.405 & 570 & 0.383 & 4.098...4.111 & 3.386...3.791 & 11.804...14.543 \\
(16232)  Chijagerbs & C10 & 2457603.7962 & 2457609.5381 & 5.742 & 278 & 0.985 & 3.702...3.707 & 3.199...3.276 & 15.381...15.785 \\
(177943) 2005 VZ17 & C10 & 2457582.5862 & 2457589.7788 & 7.193 & 343 & 0.971 & 3.532...3.539 & 2.857...2.946 & 14.443...15.366 \\
(341841) 2008 DN52 & C10 & 2457582.5862 & 2457589.7788 & 7.193 & 340 & 0.963 & 3.565...3.574 & 2.806...2.892 & 13.063...14.198 \\
(43818) 1992 ET32 & C10 & 2457582.5862 & 2457607.8830 & 25.297 & 544 & 0.439 & 3.492...3.503 & 2.723...3.029 & 13.207...16.535 \\
(90704) 1988 RO12 & C10 & 2457584.6500 & 2457605.9622 & 21.312 & 297 & 0.284 & 3.931...3.937 & 3.335...3.624 & 13.633...15.240 \\
(99281) 2001 QR99 & C10 & 2457582.5862 & 2457589.7788 & 7.193 & 339 & 0.960 & 1.872...1.877 & 1.436...1.506 & 33.695...33.872 \\
(30764) 1981 EK47 & C12 & 2457738.3719 & 2457770.6570 & 32.285 & 1531 & 0.968 & 4.476...4.485 & 3.657...4.069 & 10.005...9.970 \\
(58279) 1993 TE40 & C12 & 2457738.3719 & 2457747.5057 & 9.134 & 421 & 0.939 & 4.276...4.284 & 3.480...3.574 & 10.032...9.992 \\
(16843) 1997 XX3 & C12 & 2457743.4394 & 2457759.3572 & 15.918 & 597 & 0.765 & 3.535...3.558 & 2.794...3.013 & 12.010...14.380 \\
(16915) 1998 FR10 & C12 & 2457756.0061 & 2457771.1065 & 15.100 & 733 & 0.990 & 4.135...4.153 & 3.474...3.700 & 11.070...12.736 \\
(31338) 1998 HX147 & C12 & 2457779.7295 & 2457786.1048 & 6.375 & 274 & 0.875 & 4.193...4.195 & 3.862...3.960 & 13.131...13.364 \\
(65989) 1998 KZ12 & C12 & 2457743.6846 & 2457767.1628 & 23.478 & 1014 & 0.881 & 4.444...4.461 & 3.702...4.038 & 10.022...9.986 \\
(98002) 2000 QG199 & C12 & 2457748.7522 & 2457763.7096 & 14.957 & 705 & 0.961 & 3.923...3.944 & 3.228...3.444 & 11.336...13.179 \\
(76811) 2000 QK57 & C12 & 2457753.1658 & 2457768.0006 & 14.835 & 622 & 0.855 & 4.192...4.198 & 3.544...3.756 & 11.073...12.657 \\
(130453) 2000 QT59 & C12 & 2457763.9343 & 2457778.5239 & 14.590 & 677 & 0.946 & 3.727...3.746 & 3.209...3.439 & 13.859...14.870 \\
(77893) 2001 SM251 & C12 & 2457740.3335 & 2457754.3306 & 13.997 & 665 & 0.969 & 3.893...3.905 & 3.077...3.245 & 10.011...9.963 \\
(131502) 2001 SW273 & C12 & 2457743.7255 & 2457759.0303 & 15.305 & 703 & 0.937 & 3.717...3.739 & 2.993...3.211 & 11.645...13.769 \\
(77892) 2001 SZ250 & C12 & 2457760.8693 & 2457775.7858 & 14.917 & 714 & 0.976 & 4.515...4.523 & 3.911...4.116 & 10.492...11.907 \\
(83801) 2001 TG218 & C12 & 2457738.3719 & 2457766.8768 & 28.505 & 1223 & 0.875 & 3.978...4.027 & 3.220...3.539 & 10.019...9.971 \\
(83722) 2001 TL98 & C12 & 2457750.3869 & 2457772.2917 & 21.905 & 1036 & 0.965 & 4.441...4.451 & 3.725...4.036 & 10.009...9.973 \\
(77903) 2001 TQ142 & C12 & 2457769.6149 & 2457774.5189 & 4.904 & 218 & 0.904 & 4.543...4.544 & 4.127...4.200 & 11.773...12.051 \\
(55505) 2001 UK113 & C12 & 2457738.3719 & 2457772.0465 & 33.675 & 1596 & 0.967 & 4.472...4.473 & 3.632...4.063 & 10.020...9.985 \\
(90456) 2004 CV2 & C12 & 2457765.6099 & 2457780.5264 & 14.916 & 367 & 0.502 & 4.479...4.492 & 3.896...4.100 & 10.703...12.078 \\
(216411) 2008 RR51 & C12 & 2457741.0487 & 2457751.2655 & 10.217 & 486 & 0.969 & 3.664...3.674 & 2.832...2.950 & 10.038...9.983 \\
(7394)  Xanthomalitia & C12 & 2457746.9336 & 2457761.9931 & 15.059 & 727 & 0.984 & 3.802...3.803 & 3.108...3.307 & 11.760...13.752 \\
(193354) 2000 UX34 & C13 & 2457830.4866 & 2457850.5524 & 20.066 & 505 & 0.513 & 3.247...3.276 & 2.484...2.746 & 12.512...16.050 \\
(193449) 2000 WW146 & C13 & 2457822.1701 & 2457837.0662 & 14.896 & 673 & 0.921 & 2.670...2.692 & 1.889...2.063 & 15.147...18.503 \\
(195204) 2002 CR306 & C13 & 2457835.2272 & 2457849.8576 & 14.630 & 642 & 0.895 & 3.461...3.480 & 2.742...2.936 & 12.531...14.896 \\
(185290) 2006 UB219 & C13 & 2457831.2835 & 2457846.3022 & 15.019 & 561 & 0.762 & 3.832...3.844 & 3.183...3.368 & 11.931...13.967 \\
(403237) 2008 VE11 & C13 & 2457826.5429 & 2457841.5820 & 15.039 & 677 & 0.918 & 2.825...2.835 & 2.046...2.215 & 14.255...17.621 \\
(19034)  Santorini & C13 & 2457830.8339 & 2457845.7096 & 14.876 & 647 & 0.887 & 3.136...3.157 & 2.379...2.567 & 13.189...16.012 \\
(1748) 1951 XD & C14 & 2457929.8960 & 2457941.8905 & 11.995 & 579 & 0.984 & 4.699...4.708 & 4.131...4.283 & 11.205...12.221 \\
(13035) 1977 CE2 & C14 & 2457926.4836 & 2457936.9660 & 10.482 & 509 & 0.990 & 3.981...3.995 & 3.353...3.469 & 12.734...13.984 \\
(7174) 1988 SQ & C14 & 2457928.9356 & 2457938.0899 & 9.154 & 444 & 0.988 & 4.168...4.181 & 3.566...3.670 & 12.392...13.407 \\
(120962) 1998 VM11 & C14 & 2457925.2167 & 2457937.2725 & 12.056 & 560 & 0.947 & 4.635...4.640 & 4.038...4.205 & 11.200...12.274 \\
(134652) 1999 VT37 & C14 & 2457923.4186 & 2457935.5357 & 12.117 & 514 & 0.865 & 4.586...4.603 & 3.981...4.159 & 11.250...12.328 \\
(60381) 2000 AX180 & C14 & 2457924.1542 & 2457935.0044 & 10.850 & 515 & 0.967 & 4.401...4.409 & 3.737...3.865 & 11.097...12.387 \\
(86435) 2000 CL9 & C14 & 2457929.3034 & 2457939.7654 & 10.462 & 488 & 0.951 & 4.227...4.236 & 3.605...3.727 & 12.073...13.205 \\
(18916) 2000 OG44 & C14 & 2457930.1821 & 2457946.6516 & 16.470 & 670 & 0.830 & 4.196...4.271 & 3.581...3.874 & 12.359...13.626 \\
(208290) 2001 DH1 & C14 & 2457920.7622 & 2457927.6892 & 6.927 & 324 & 0.952 & 4.056...4.064 & 3.419...3.496 & 12.413...13.284 \\
(63293) 2001 DT89 & C14 & 2457923.6025 & 2457935.6992 & 12.097 & 554 & 0.934 & 4.132...4.147 & 3.516...3.657 & 12.305...13.657 \\
(174074) 2002 EO133 & C14 & 2457925.9932 & 2457938.0082 & 12.015 & 518 & 0.879 & 4.285...4.304 & 3.697...3.838 & 12.060...13.300 \\
(174077) 2002 FP21 & C14 & 2457930.1208 & 2457941.4819 & 11.361 & 541 & 0.971 & 4.138...4.151 & 3.565...3.702 & 12.767...13.878 \\
(146961) 2002 GH129 & C14 & 2457921.4978 & 2457933.5945 & 12.097 & 406 & 0.684 & 3.765...3.780 & 3.101...3.234 & 13.053...14.688 \\
(119942) 2002 GJ129 & C14 & 2457925.1963 & 2457937.2725 & 12.076 & 513 & 0.866 & 4.229...4.239 & 3.609...3.755 & 12.039...13.338 \\
(174089) 2002 GX137 & C14 & 2457927.7505 & 2457939.7859 & 12.035 & 547 & 0.927 & 3.641...3.657 & 2.962...3.089 & 13.390...15.118 \\
(141557) 2002 GY69 & C14 & 2457923.3981 & 2457935.3518 & 11.954 & 504 & 0.859 & 3.743...3.754 & 3.065...3.195 & 13.032...14.665 \\
(177640) 2003 JF14 & C14 & 2457925.9728 & 2457938.0899 & 12.117 & 579 & 0.974 & 4.637...4.651 & 4.079...4.230 & 11.360...12.410 \\
(117113) 2004 PG11 & C14 & 2457931.0199 & 2457943.0553 & 12.035 & 397 & 0.672 & 4.294...4.301 & 3.676...3.826 & 11.945...13.189 \\
(207644) 2006 UW322 & C14 & 2457923.4594 & 2457935.5561 & 12.097 & 557 & 0.939 & 3.734...3.734 & 3.029...3.172 & 12.901...14.587 \\
(39415)  Janeausten & C14 & 2457929.6304 & 2457936.3122 & 6.682 & 323 & 0.984 & 4.189...4.199 & 3.651...3.728 & 12.899...13.540 \\
(90704) 1988 RO12 & C15 & 2458005.4390 & 2458020.5395 & 15.101 & 612 & 0.826 & 4.031...4.034 & 3.301...3.490 & 11.814...13.621 \\
(11274) 1988 SX2 & C15 & 2458007.7480 & 2458021.5611 & 13.813 & 502 & 0.741 & 4.805...4.806 & 4.181...4.370 & 10.754...11.866 \\
(7284) 1989 VW & C15 & 2458010.0979 & 2458024.9940 & 14.896 & 688 & 0.942 & 4.581...4.588 & 3.915...4.118 & 10.924...12.274 \\
(178295) 1992 DJ6 & C15 & 2458008.9127 & 2458024.0336 & 15.121 & 595 & 0.802 & 4.214...4.232 & 3.498...3.705 & 11.406...13.038 \\
(152900) 2000 DC27 & C15 & 2457999.4724 & 2458014.6137 & 15.141 & 608 & 0.819 & 4.112...4.131 & 3.387...3.592 & 11.642...13.352 \\
(92283) 2000 DC45 & C15 & 2457997.7151 & 2458012.7134 & 14.998 & 462 & 0.628 & 3.739...3.746 & 2.970...3.150 & 12.290...14.360 \\
(29053) 2000 DL95 & C15 & 2458006.1338 & 2458021.1525 & 15.019 & 656 & 0.891 & 4.045...4.053 & 3.315...3.509 & 11.773...13.555 \\
(176158) 2001 HG21 & C15 & 2457998.6755 & 2458013.7759 & 15.100 & 692 & 0.935 & 3.677...3.686 & 2.903...3.086 & 12.426...14.571 \\
(114954) 2003 QE57 & C15 & 2458008.5858 & 2458023.5636 & 14.978 & 564 & 0.768 & 4.295...4.327 & 3.675...3.840 & 11.788...13.260 \\
(147836) 2005 TN125 & C15 & 2458000.3919 & 2458015.4719 & 15.080 & 567 & 0.767 & 3.889...3.908 & 3.137...3.337 & 12.009...13.911 \\
(161606) 2005 UR38 & C15 & 2458001.6792 & 2458016.7388 & 15.060 & 594 & 0.804 & 4.215...4.217 & 3.515...3.705 & 11.600...13.241 \\
(249182) 2008 CW119 & C15 & 2458004.1721 & 2458019.1704 & 14.998 & 642 & 0.873 & 3.712...3.723 & 2.946...3.130 & 12.398...14.448 \\
(62489) 2000 SS223 & C16 & 2458157.8941 & 2458167.1505 & 9.256 & 330 & 0.726 & 3.210...3.217 & 2.494...2.596 & 13.328...15.148 \\
(87956) 2000 TM4 & C16 & 2458149.9046 & 2458160.8366 & 10.932 & 494 & 0.921 & 3.175...3.189 & 2.487...2.603 & 13.868...15.889 \\
(194512) 2001 XR & C16 & 2458148.9646 & 2458159.3449 & 10.380 & 311 & 0.611 & 3.479...3.495 & 2.838...2.955 & 13.205...14.828 \\
(1748) 1951 XD & C17 & 2458200.7434 & 2458210.8171 & 10.074 & 475 & 0.961 & 4.386...4.401 & 4.054...4.226 & 12.640...12.867 \\
(233980) 1995 NF1 & C17 & 2458219.7875 & 2458240.8341 & 21.047 & 911 & 0.883 & 3.824...3.866 & 3.074...3.387 & 11.105...13.751 \\
(120962) 1998 VM11 & C17 & 2458208.3242 & 2458229.3300 & 21.006 & 844 & 0.820 & 4.647...4.653 & 3.986...4.297 & 10.015...9.984 \\
(119918) 2002 EX84 & C17 & 2458237.7486 & 2458246.5759 & 8.827 & 420 & 0.969 & 3.609...3.621 & 2.797...2.905 & 10.694...12.346 \\
(111995) 2002 GG114 & C17 & 2458207.2821 & 2458220.9522 & 13.670 & 610 & 0.910 & 4.363...4.372 & 3.909...4.124 & 12.237...12.884 \\
(117106) 2004 OT6 & C17 & 2458219.7875 & 2458240.8954 & 21.108 & 722 & 0.698 & 4.099...4.112 & 3.375...3.670 & 10.676...13.100 \\
(236209) 2005 WT122 & C17 & 2458217.8463 & 2458238.9133 & 21.067 & 936 & 0.906 & 4.087...4.110 & 3.369...3.677 & 10.775...13.139 \\
(207638) 2006 TJ32 & C17 & 2458217.2741 & 2458238.3616 & 21.087 & 808 & 0.782 & 3.930...3.942 & 3.194...3.484 & 10.968...13.586 \\
(237323) 2009 BP88 & C17 & 2458226.8575 & 2458245.7382 & 18.881 & 796 & 0.860 & 3.835...3.850 & 2.985...3.209 & 10.010...9.960 \\
(58353) 1995 EW4 & C18 & 2458253.3190 & 2458273.8548 & 20.536 & 879 & 0.873 & 3.772...3.803 & 2.993...3.261 & 10.942...14.121 \\
(193241) 2000 ST33 & C18 & 2458255.3011 & 2458276.3068 & 21.006 & 963 & 0.935 & 3.448...3.498 & 2.685...2.979 & 12.485...15.691 \\
(36941) 2000 SV239 & C18 & 2458257.7531 & 2458278.7180 & 20.965 & 923 & 0.898 & 3.661...3.682 & 2.919...3.189 & 12.068...15.054 \\
(57027) 2000 UB59 & C18 & 2458256.6497 & 2458277.7167 & 21.067 & 928 & 0.899 & 3.655...3.677 & 2.912...3.187 & 12.068...15.083 \\
(45862) 2000 UQ51 & C18 & 2458282.3756 & 2458297.8030 & 15.427 & 706 & 0.933 & 4.080...4.101 & 3.585...3.818 & 13.492...14.446 \\
(270356) 2001 YU49 & C18 & 2458251.5413 & 2458270.8306 & 19.289 & 861 & 0.911 & 3.143...3.171 & 2.379...2.621 & 13.804...17.114 \\
(230294) 2001 YW79 & C18 & 2458258.3252 & 2458279.2697 & 20.945 & 981 & 0.956 & 3.700...3.732 & 2.965...3.248 & 12.038...14.896 \\
(274543) 2008 SJ248 & C18 & 2458259.1221 & 2458277.2468 & 18.125 & 789 & 0.888 & 3.316...3.339 & 2.554...2.779 & 13.147...16.150 \\
(2483)  Guinevere & C18 & 2458251.5413 & 2458270.4219 & 18.881 & 846 & 0.914 & 3.231...3.272 & 2.478...2.731 & 13.598...16.610 \\
(15278)  Paquet & C18 & 2458259.6330 & 2458275.6121 & 15.979 & 718 & 0.916 & 3.792...3.822 & 3.095...3.318 & 12.297...14.344 \\
\enddata
\end{deluxetable*}


\startlongtable
\begin{deluxetable*}{lccccll}
\tablecaption{Periods of K2 Hildas, ordered by the asteroid numbers. The columns are: (1) the number of the asteroid; (2) the campaign of the observation; (3) observation in SDSS, the number of detections in parenthesis; (4) sinodical rotation period in days; (5) peak-to-peak amplitude in magnitude; (6) period and amplitude in previous publications, and the reference; (7) remarks. For Hildas which were detected in multiple campaigns we derived the period and the amplitude for each campaigns separately.}
\tablehead{\colhead{Number} & \colhead{Campaign} & \colhead{SDSS} & \colhead{Period (d)} & \colhead{Amplitude (mag)} & \colhead{Values in prev. publ.}  & \colhead{Remarks}}
\colnumbers
\startdata
1748    &  C14  &  yes (1)  &  6.005    &  0.09  &  6.00, 0.12 (Dahlgren 1998)    &  three humps \\
        &  C17  &  yes (1)  &  6.000    &  0.07  &  6.001, 0.1 (Slyusarev 2012)   &  three humps \\
        &       &           &           &        &  5.552, 0.25 (Warner 2017c)    & \\
        &       &           &           &        &  5.320, 0.08 (Warner 2018c)    & \\
        &       &           &           &        &  5.551, 0.23 (Warner 2018c)    & \\
2483    &  C18  &  yes (2)  &  14.733   &  1.38  &  14.733, 1.38 (Dahlgren 1998)  & \\
        &       &           &           &        &  14.731, 1.37 (Durech 2016)    & \\
        &       &           &           &        &  14.730, 0.89 (Warner 2017b)   & \\
        &       &           &           &        &  14.721, 1.37 (Warner 2018a)   & \\
3655    &  C06  &  yes (1)  &  77.047   &  0.19  &  \textgreater20, 0.07 (Dahlgren 1998)  &  three humps \\
        &       &           &           &        &  75.668, 0.23 (Waszczak 2015)  & \\
7174    &  C14  &  no       &  34.433   &  0.13  &  7.456, 0.38 (Warner 2017c)    & \\
7284    &  C10  &  no       &  26.17    &  0.08  &                                & \\
        &  C15  &  no       &  26.42    &  0.08  &                                & \\
7394    &  C12  &  yes (3)  &  4.836    &  0.16  &                                & \\
8550    &  C06  &  yes (2)  &  37.209   &  0.28  &  6.719, 0.09 (Warner 2019)    & \\
11274   &  C15  &  no       &  158.940  &  0.86  &                                & \\
13035   &  C14  &  yes      &  10.660   &  0.44  &  10.657, --- (Durech 2016)     & \\
        &       &           &           &        &  10.639, 0.57 (Warner 2018c)   & \\
15278   &  C18  &  yes (1)  &  39.160   &  0.26  &  40.01, 0.41 (Warner 2017a)    & \\
15626   &  C06  &  yes (9)  &  13.544   &  0.75  &  13.277, 0.68 (Waszczak 2015)  & \\
16232   &  C10  &  yes (1)  &  152.400  &  0.69  &                                & \\
16843   &  C12  &  yes (2)  &  309.5    &  0.28  &  275, 0.41 (Warner 2017a)      & \\
16915   &  C12  &  no       &  425.155  &  0.95  &                                &  very long per. \\
18916   &  C14  &  no       &  22.493   &  0.18  &                                & \\
19034   &  C13  &  no       &  280.784  &  0.79  &  247, 0.43 (Warner 2017b)      & \\
20628   &  C08  &  yes (2)  &  320.0    &  0.95  &  68.1, 1.04 (Warner 2018b)     &  tumbler \\
20630   &  C08  &  no       &  25.750   &  0.43  &                                & \\
27561   &  C11  &  yes (1)  &  ---      &  \textless0.1  &                        &  no signal \\
29053   &  C15  &  no       &  17.980   &  0.24  &                                & \\
30764   &  C12  &  yes (2)  &  24.930   &  0.41  &                                & \\
31338   &  C12  &  no       &  7.670    &  0.18  &                                & \\
36941   &  C18  &  no       &  36.954   &  0.21  &                                & \\
39301   &  C06  &  no       &  12.572   &  0.42  &                                & \\
39415   &  C14  &  no       &  7.317    &  0.62  &                                & \\
43818   &  C10  &  yes (3)  &  21.640   &  0.29  &                                & \\
45862   &  C18  &  yes (1)  &  13.159   &  1.00  &                                & \\
46302   &  C06  &  yes (1)  &  7.479    &  0.14  &                                & \\
51930   &  C06  &  yes (1)  &  52.288   &  0.37  &                                & \\
55505   &  C12  &  no       &  4.460    &  0.25  &                                & \\
57027   &  C18  &  yes (4)  &  4.800    &  0.15  &                                &  rotation \\
        &       &           &  25.798   &  0.12  &                                &  orbit \\
58279   &  C12  &  no       &  6.836    &  0.18  &                                & \\
58353   &  C18  &  no       &  156.550  &  0.53  &                                & \\
60381   &  C14  &  yes (6)  &  5.897    &  0.17  &  28.96, 0.30 (Warner 2018c)    &  three humps \\
62489   &  C16  &  yes (1)  &  63.583   &  0.58  &                                & \\
63293   &  C14  &  no       &  199.68   &  0.53  &                                & \\
65989   &  C12  &  yes (1)  &  35.357   &  0.05  &                                & \\
76811   &  C12  &  yes (2)  &  26.754   &  0.15  &                                & \\
77892   &  C12  &  no       &  26.388   &  0.26  &                                & \\
77893   &  C12  &  no       &  6.028    &  0.09  &                                & \\
78159   &  C06  &  yes (5)  &  15.404   &  0.22  &                                & \\
83722   &  C12  &  yes (1)  &  9.074    &  0.65  &                                & \\
83801   &  C12  &  no       &  18.721   &  0.05  &                                & \\
86435   &  C14  &  no       &  47.0     &  0.08  &                                & \\
87956   &  C16  &  yes (2)  &  160.6    &  0.49  &                                & \\
90456   &  C12  &  yes (3)  &  17.622   &  0.12  &                                & \\
90704 &  C10  &  yes (1)  &  4.760    &  0.15  &                                &  tumbler P1 \\
        &       &           &  23.762   &  0.27  &                                &  tumbler P2 \\
        &  C15  &  yes (1)  &  4.735    &  0.16  &                                & \\
92281   &  C10  &  yes (4)  &  65.529   &  0.09  &                                & \\
92283   &  C15  &  no       &  5.490    &  0.29  &                                & \\
92284   &  C10  &  no       &  6.800    &  0.10  &                                & \\
98002        &  C12  &  yes (1)  &  7.557    &  0.25  &                                & \\
99276   &  C06  &  yes (2)  &  5.239    &  0.07  &                                & \\
111995  &  C17  &  yes (1)  &  ---      &  ---   &                                &  no signal \\
113224  &  C06  &  no       &  8.624    &  0.35  &                                & \\
114954  &  C15  &  no       &  39.216   &  0.32  &                                & \\
117106  &  C17  &  yes (1)  &  32.090   &  0.21  &                                & \\
117113  &  C14  &  yes (6)  &  136.908  &  0.40  &                                & \\
119918  &  C17  &  yes (8)  &  5.006    &  0.45  &                                & \\
119942  &  C14  &  no       &  5.120    &  0.52  &  5.119, 0.51 (Waszczak 2015)   & \\
120962  &  C14  &  no       &  37.134   &  0.16  &                                & \\
        &  C17  &  no       &  34.593   &  0.18  &                                & \\
130453  &  C12  &  yes (1)  &  230.7    &  0.36  &                                & \\
131502  &  C12  &  yes (1)  &  195.4    &  0.66  &                                & \\
134652  &  C14  &  no       &  33.932   &  0.21  &                                & \\
141557  &  C14  &  yes (1)  &  9.636    &  1.02  &                                & \\
146961  &  C14  &  no       &  243.7    &  0.66  &                                & \\
147836  &  C15  &  no       &  240.0    &  0.73  &                                & \\
148227  &  C10  &  no       &  6.795    &  0.03  &                                &  small amplitude \\
152133  &  C10  &  no       &  5.174    &  0.27  &                                & \\
152900  &  C15  &  no       &  10.530   &  0.77  &                                & \\
161606  &  C15  &  yes (1)  &  42.590   &  0.21  &                                & \\
174074  &  C14  &  no       &  43.015   &  0.16  &                                & \\
174077  &  C14  &  yes (1)  &  3.406    &  0.13  &                                &  tumbler P1 \\
        &  C14  &  yes (1)  &  24.328   &  0.17  &                                &  tumbler P2 \\
174089  &  C14  &  yes (5)  &  148.19   &  0.86  &                                & \\
176158  &  C15  &  no       &  194.4    &  0.62  &                                & \\
177640  &  C14  &  no       &  11.939   &  0.20  &                                & \\
177943  &  C10  &  yes (6)  &  19.020   &  0.27  &                                & \\
178295  &  C15  &  no       &  6.581    &  0.19  &                                & \\
185290 &  C08  &  no       &  85.318   &  0.15  &                                &  three humps \\
       &  C08  &  no       &  56.878   &  0.15  &                                &  dominant period \\
        &  C13  &  no       &  57.554   &  0.09  &                                & \\
193241  &  C18  &  yes (1)  &  4.203    &  0.06  &                                & \\
193354  &  C13  &  yes (1)  &  27.6     &  0.08  &                                & \\
193449  &  C13  &  yes (3)  &  21.111   &  0.20  &  38.44, 0.20 (Warner 2017b)    & \\
194512  &  C16  &  yes (1)  &  76.286   &  0.30  &                                &  \\
195204  &  C13  &  yes (1)  &  9.326    &  0.18  &                                & \\
203157  &  C08  &  no       &  6.895    &  0.79  &                                & \\
207638  &  C17  &  no       &  37.354   &  0.28  &                                & \\
207644  &  C14  &  yes (5)  &  2.689    &  0.23  &                                & \\
208290  &  C14  &  no       &  3.643    &  0.36  &                                & \\
216411  &  C12  &  yes (4)  &  22.555   &  0.95  &                                & \\
230294  &  C18  &  yes (2)  &  11.440   &  0.57  &                                & \\
233980  &  C17  &  no       &  20.739   &  0.32  &                                & \\
236209  &  C17  &  yes (11) &  6.095    &  0.15  &                                &  tumbler P1 \\
        &  C17  &  yes (11) &  18.48    &  0.15  &                                &  tumbler P2 \\
237323  &  C17  &  no       &  5.375    &  1.08  &                                & \\
249182  &  C15  &  no       &  9.364    &  0.09  &                                &  three humps \\
270356  &  C18  &  yes (1)  &  20.293   &  0.16  &                                & \\
274543  &  C18  &  no       &  432.2    &  0.81  &                                &  very long per. \\
341841  &  C10  &  yes (5)  &  4.030    &  0.18  &                                & \\
402869  &  C08  &  no       &  7.260    &  0.82  &                                & \\
403237  &  C13  &  no       &  5.597    &  0.24  &                                & \\%
\enddata
\end{deluxetable*}

\hfill
\pagebreak

\pagebreak

\newpage 

\figsetstart
\figsetnum{6}
\figsettitle{}

\figsetgrpstart
\figsetgrpnum{6.1}
\figsetgrptitle{Hilda steroid light curves, continued.}
\figsetplot{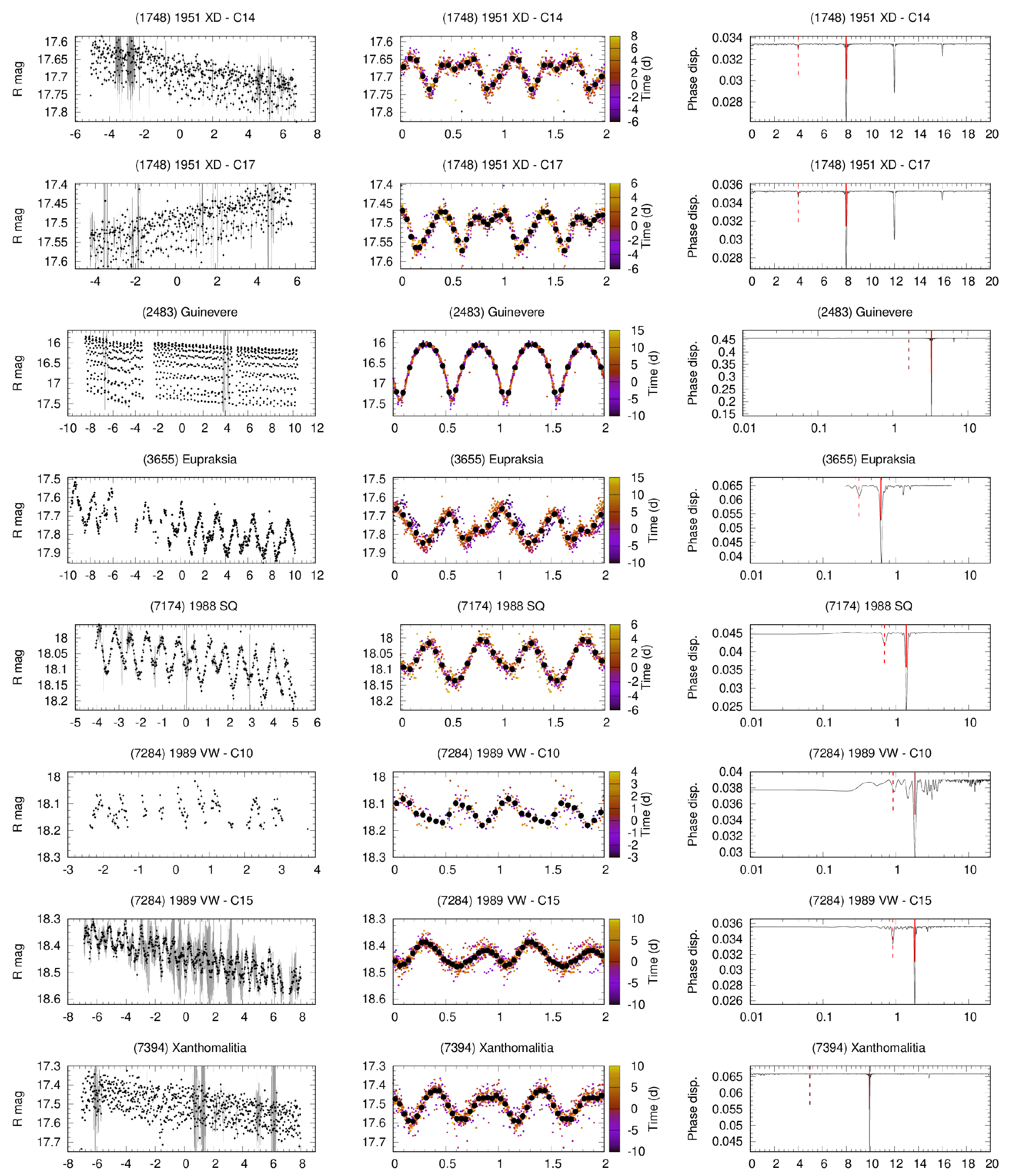}
\figsetgrpnote{Hilda asteroid light curves, continued.}
\figsetgrpend

\figsetgrpstart
\figsetgrpnum{6.2}
\figsetgrptitle{Hilda steroid light curves, continued.}
\figsetplot{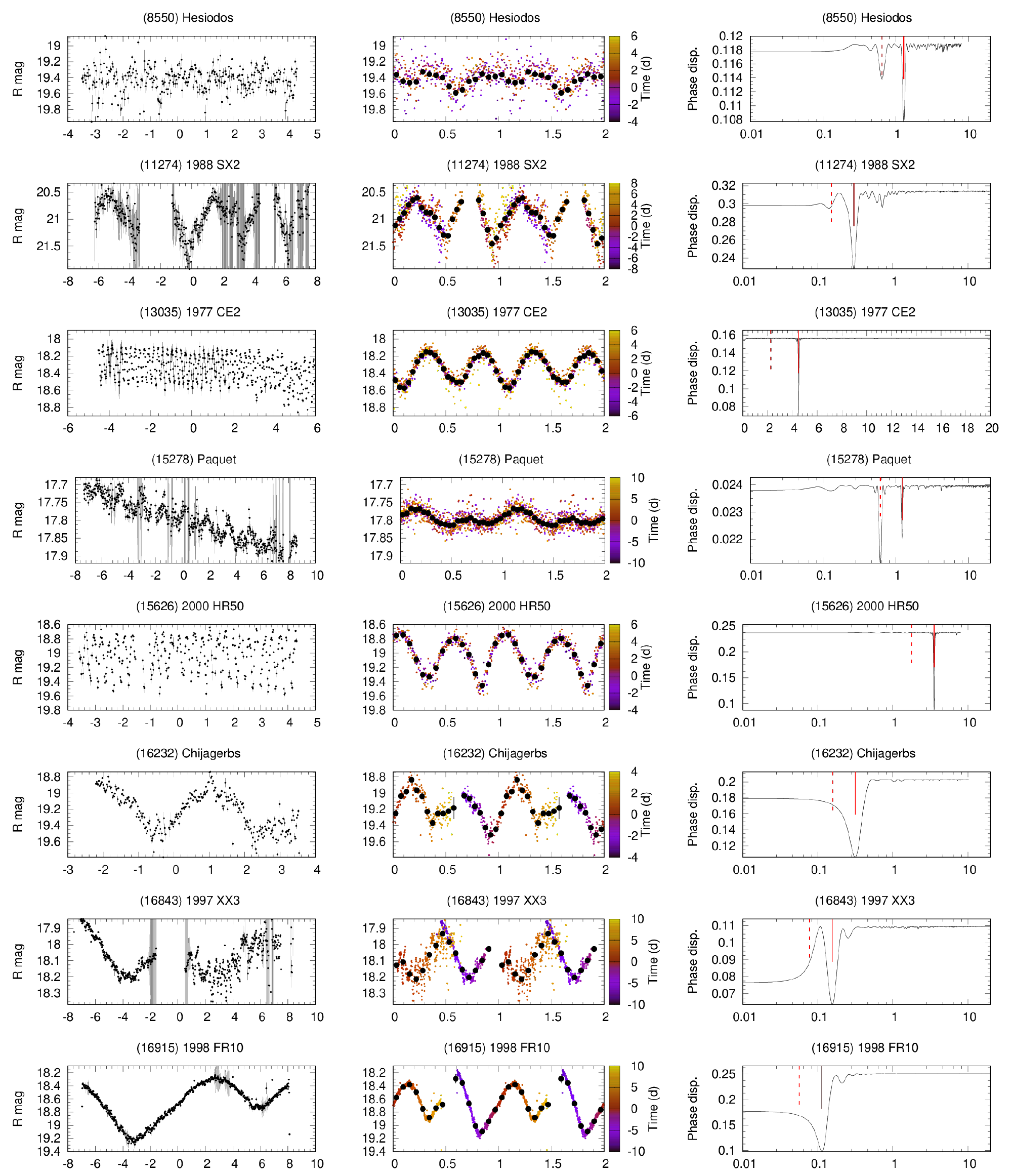}
\figsetgrpnote{Hilda asteroid light curves, continued.}
\figsetgrpend

\figsetgrpstart
\figsetgrpnum{6.3}
\figsetgrptitle{Hilda steroid light curves, continued.}
\figsetplot{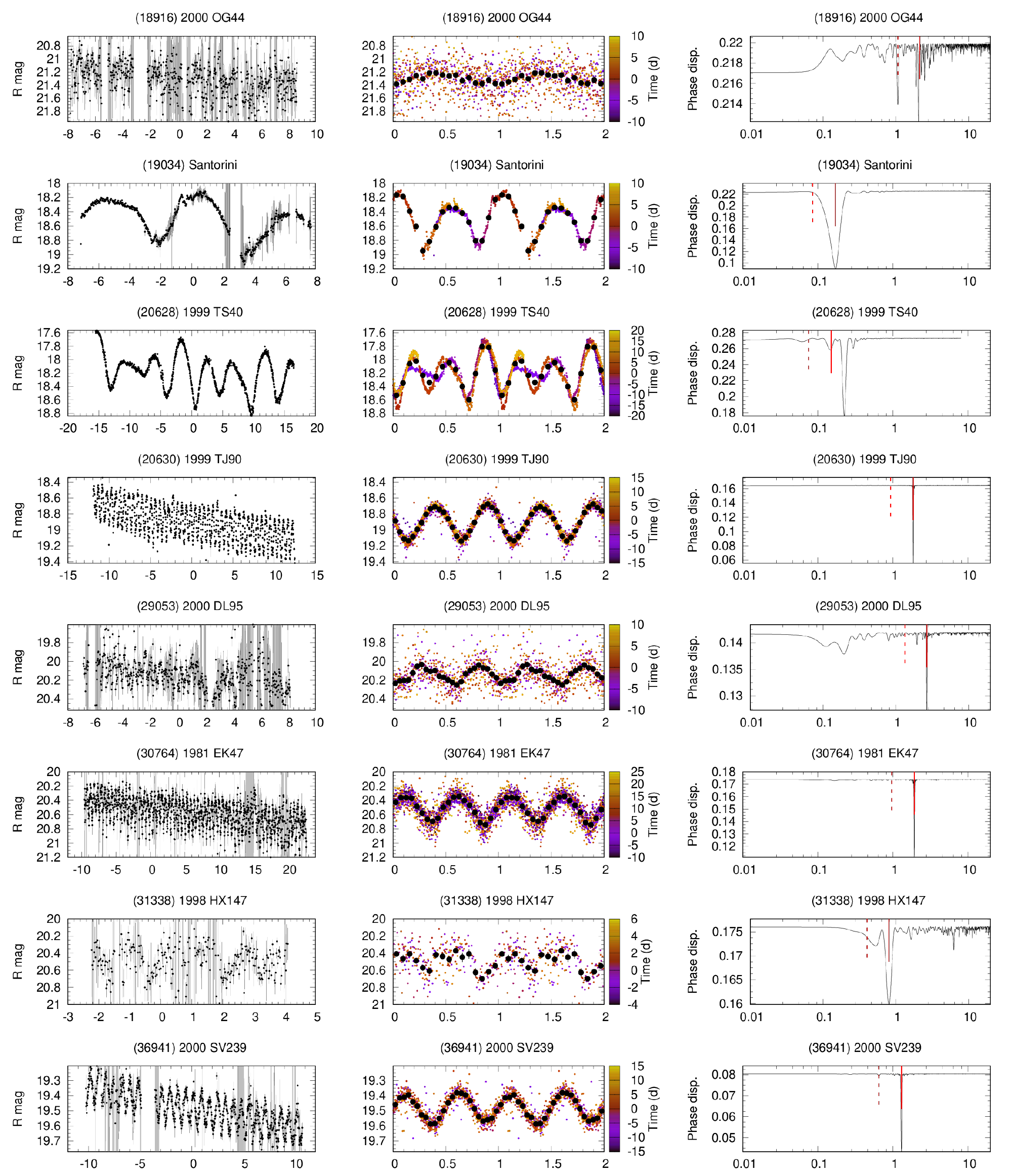}
\figsetgrpnote{Hilda asteroid light curves, continued.}
\figsetgrpend

\figsetgrpstart
\figsetgrpnum{6.4}
\figsetgrptitle{Hilda steroid light curves, continued.}
\figsetplot{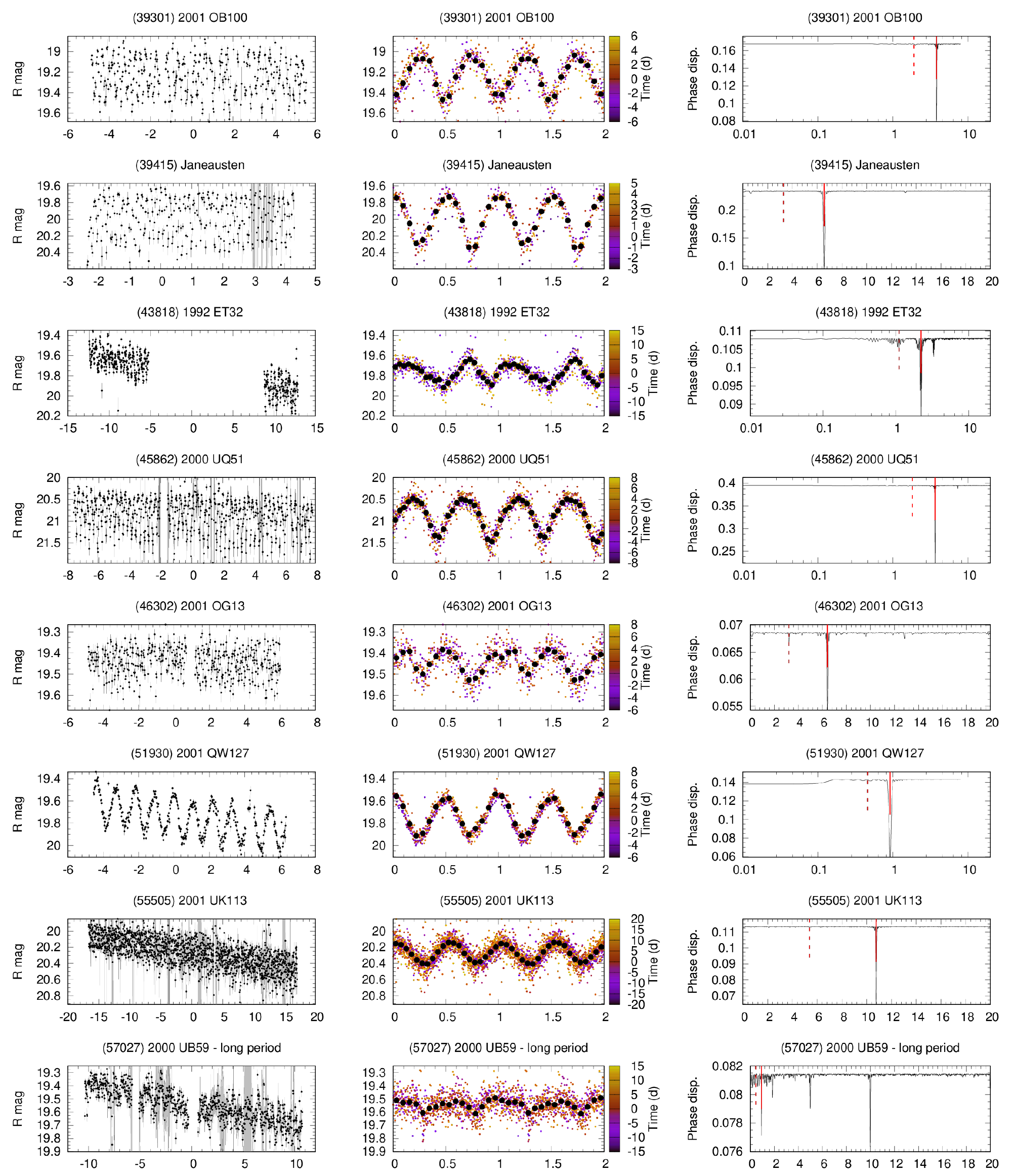}
\figsetgrpnote{Hilda asteroid light curves, continued.}
\figsetgrpend

\figsetgrpstart
\figsetgrpnum{6.5}
\figsetgrptitle{Hilda steroid light curves, continued.}
\figsetplot{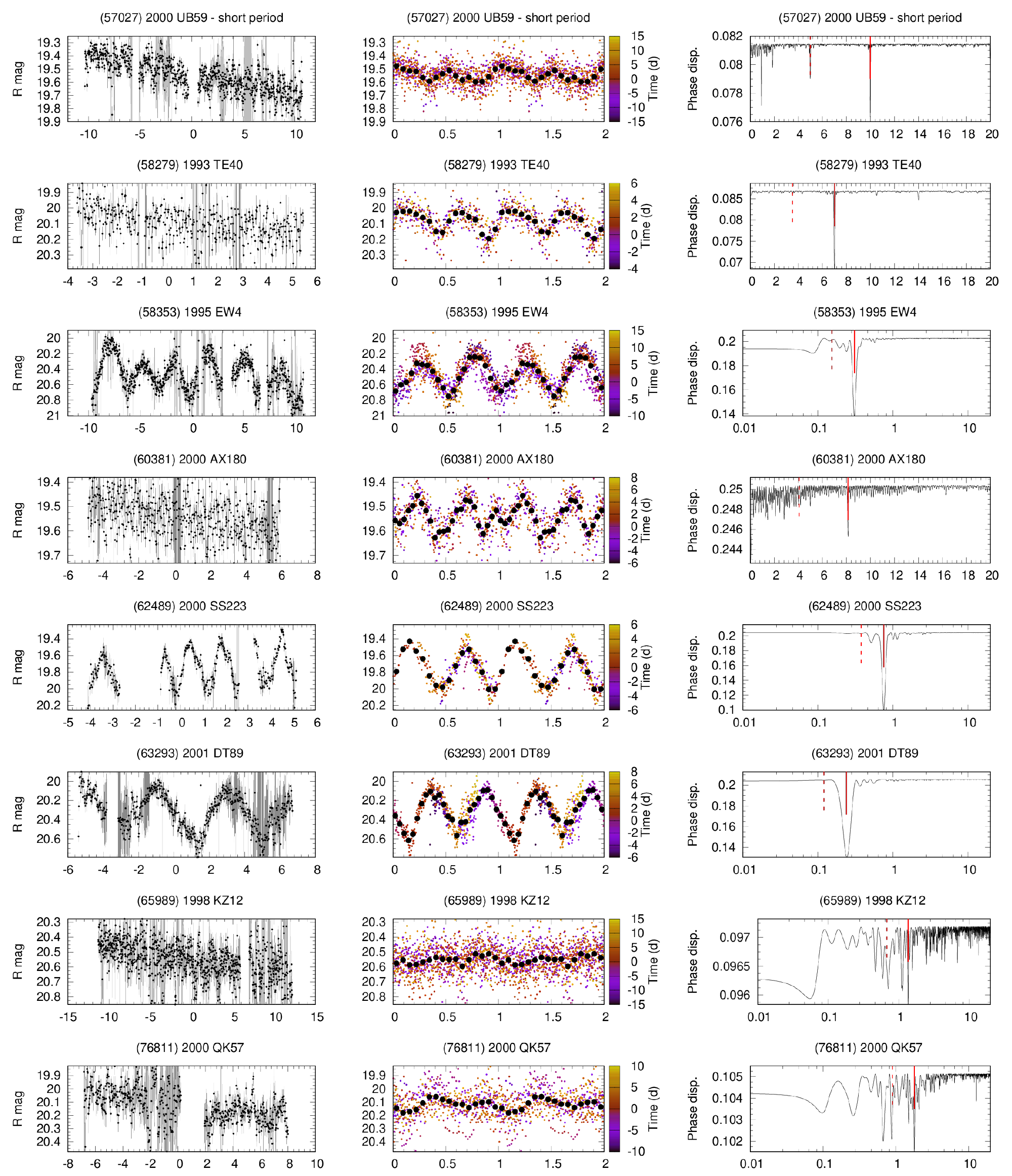}
\figsetgrpnote{Hilda asteroid light curves, continued.}
\figsetgrpend

\figsetgrpstart
\figsetgrpnum{6.6}
\figsetgrptitle{Hilda steroid light curves, continued.}
\figsetplot{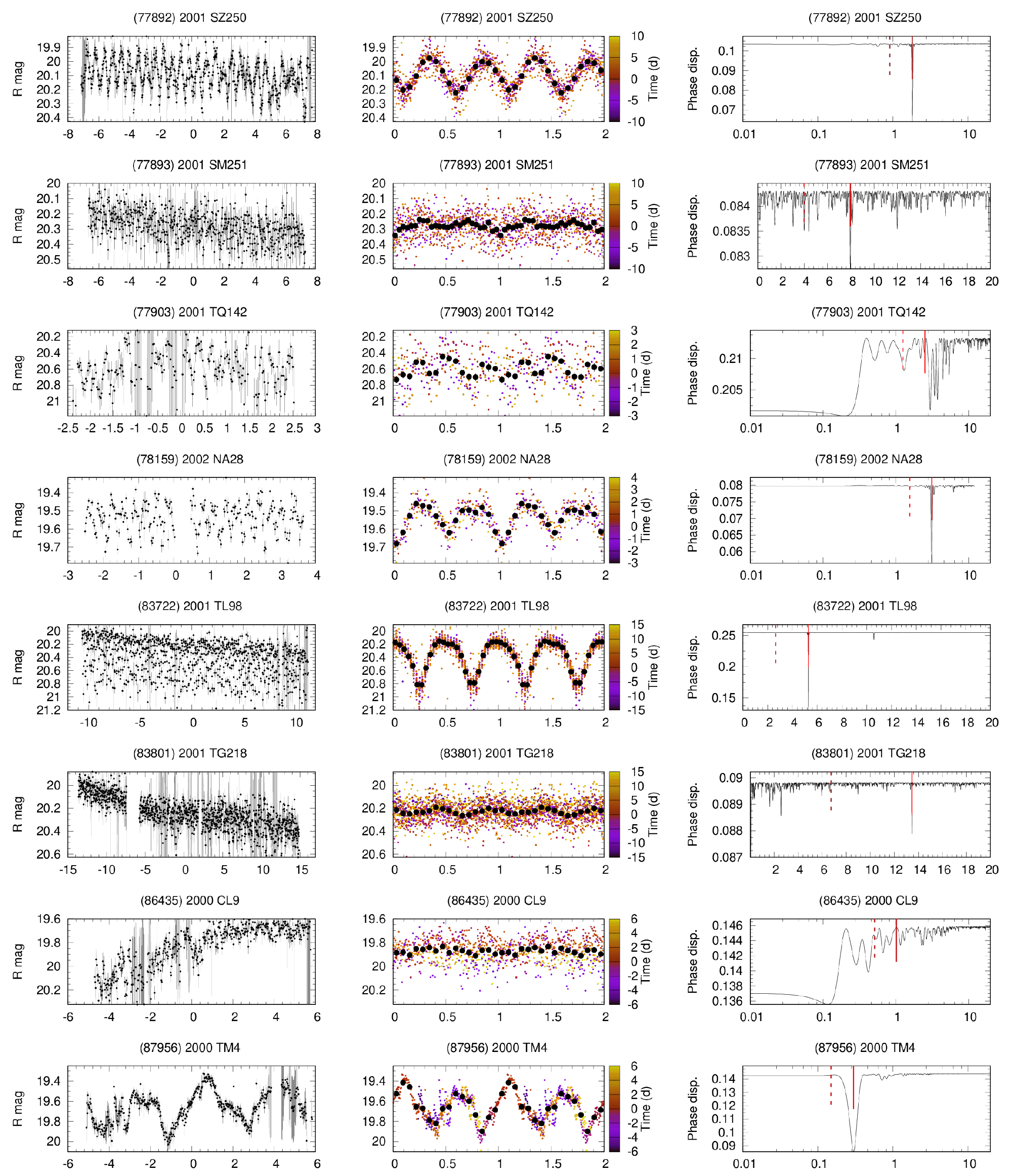}
\figsetgrpnote{Hilda asteroid light curves, continued.}
\figsetgrpend

\figsetgrpstart
\figsetgrpnum{6.7}
\figsetgrptitle{Hilda steroid light curves, continued.}
\figsetplot{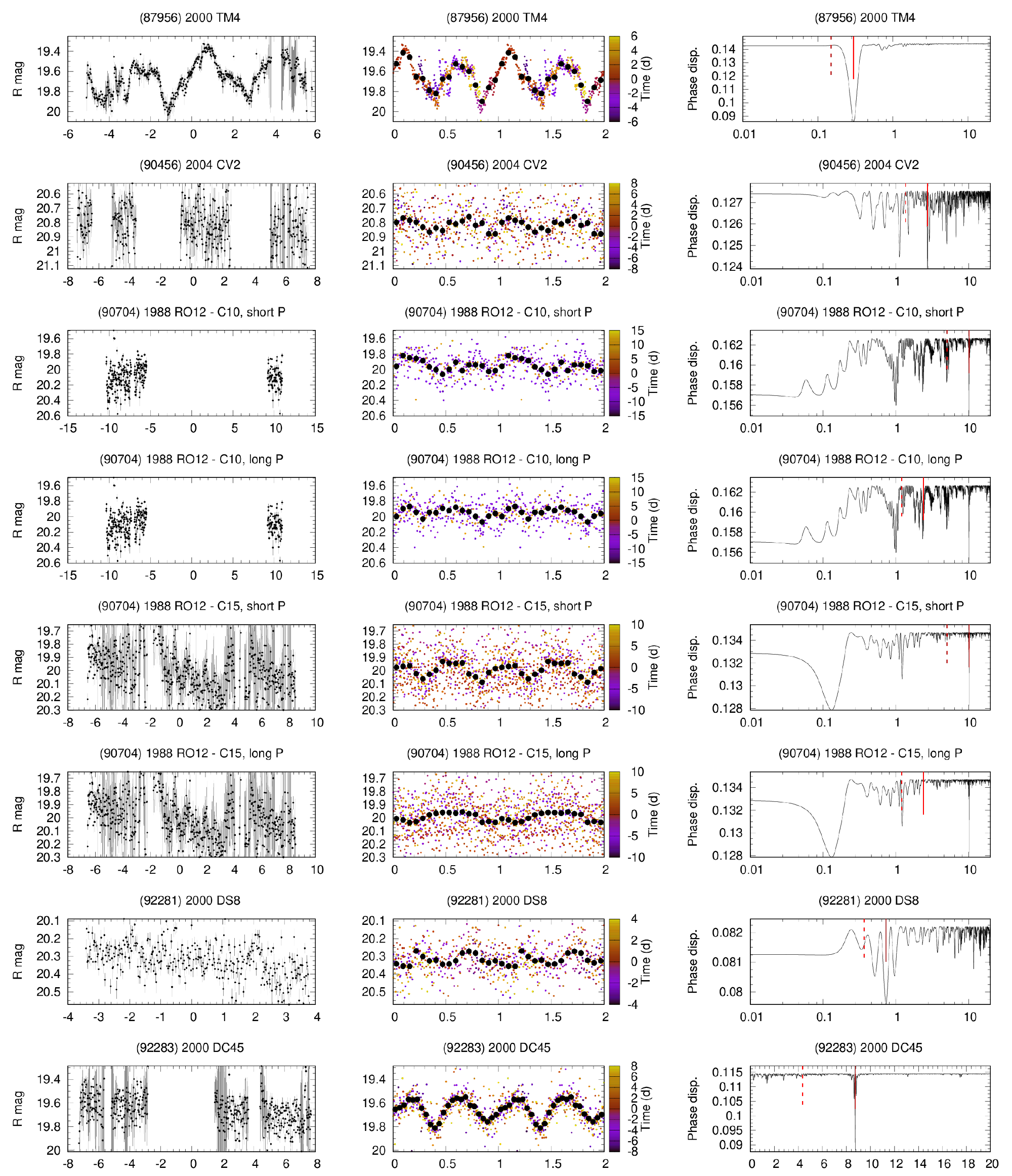}
\figsetgrpnote{Hilda asteroid light curves, continued.}
\figsetgrpend

\figsetgrpstart
\figsetgrpnum{6.8}
\figsetgrptitle{Hilda steroid light curves, continued.}
\figsetplot{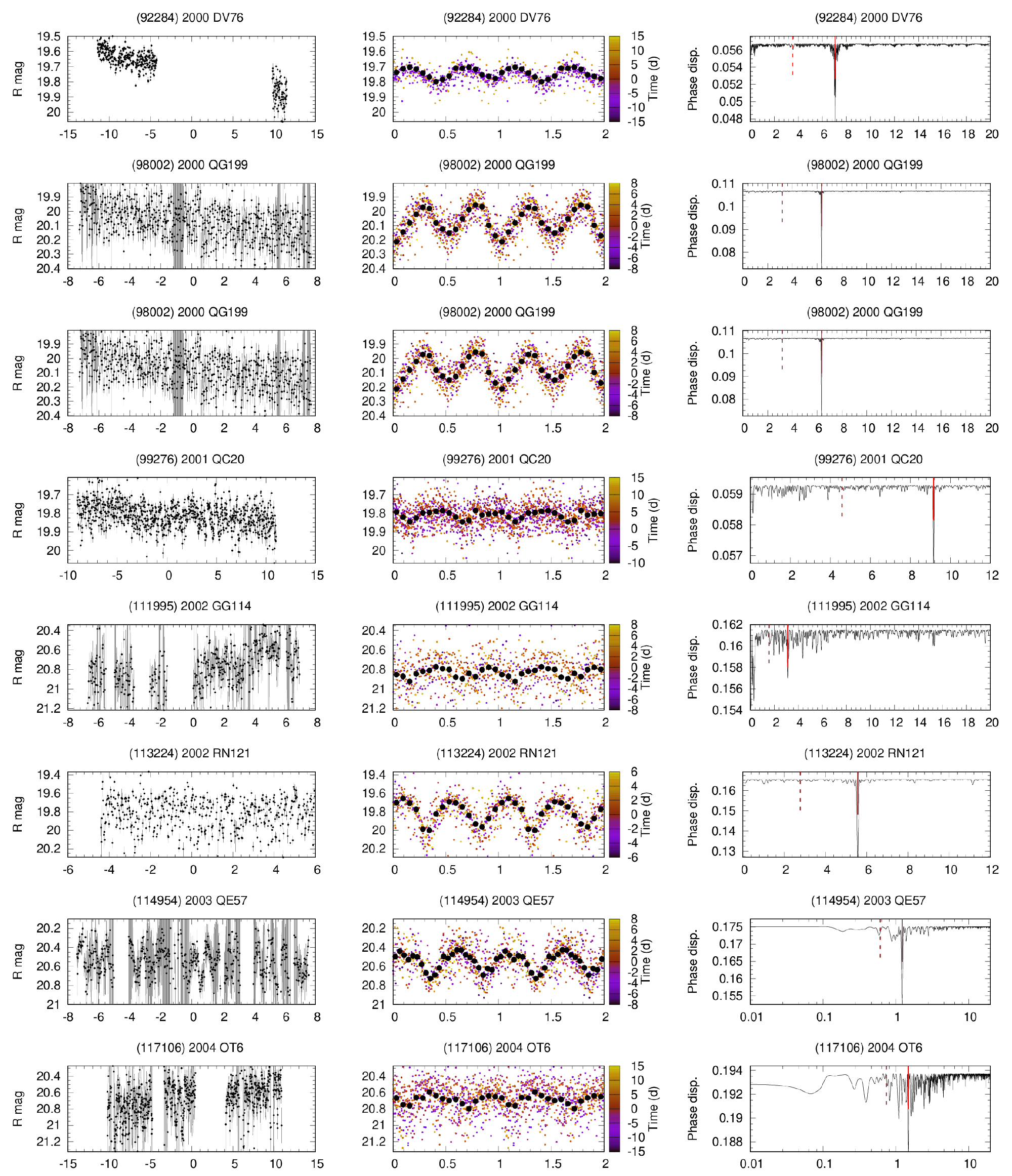}
\figsetgrpnote{Hilda asteroid light curves, continued.}
\figsetgrpend

\figsetgrpstart
\figsetgrpnum{6.9}
\figsetgrptitle{Hilda steroid light curves, continued.}
\figsetplot{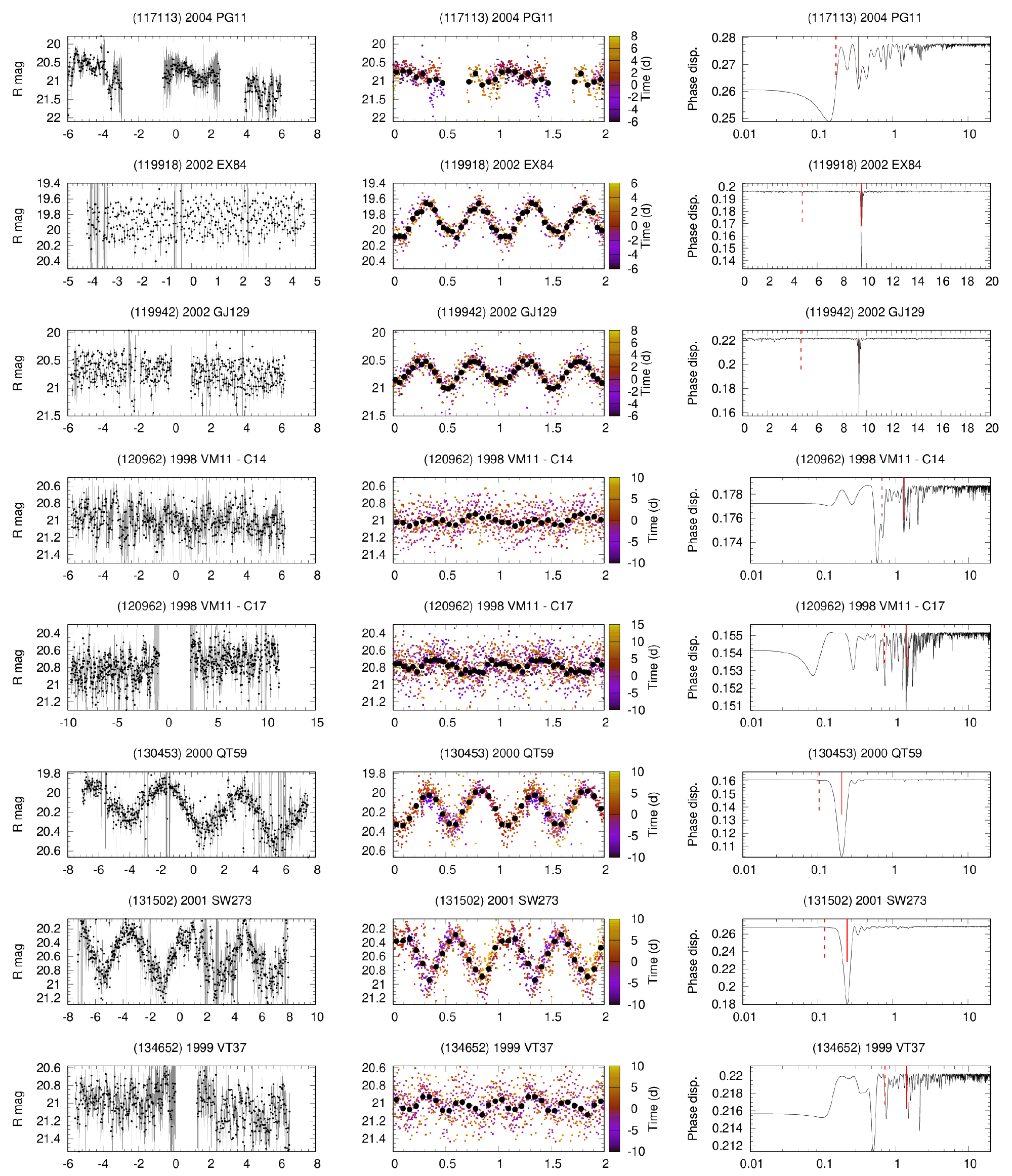}
\figsetgrpnote{Hilda asteroid light curves, continued.}
\figsetgrpend

\figsetgrpstart
\figsetgrpnum{6.10}
\figsetgrptitle{Hilda steroid light curves, continued.}
\figsetplot{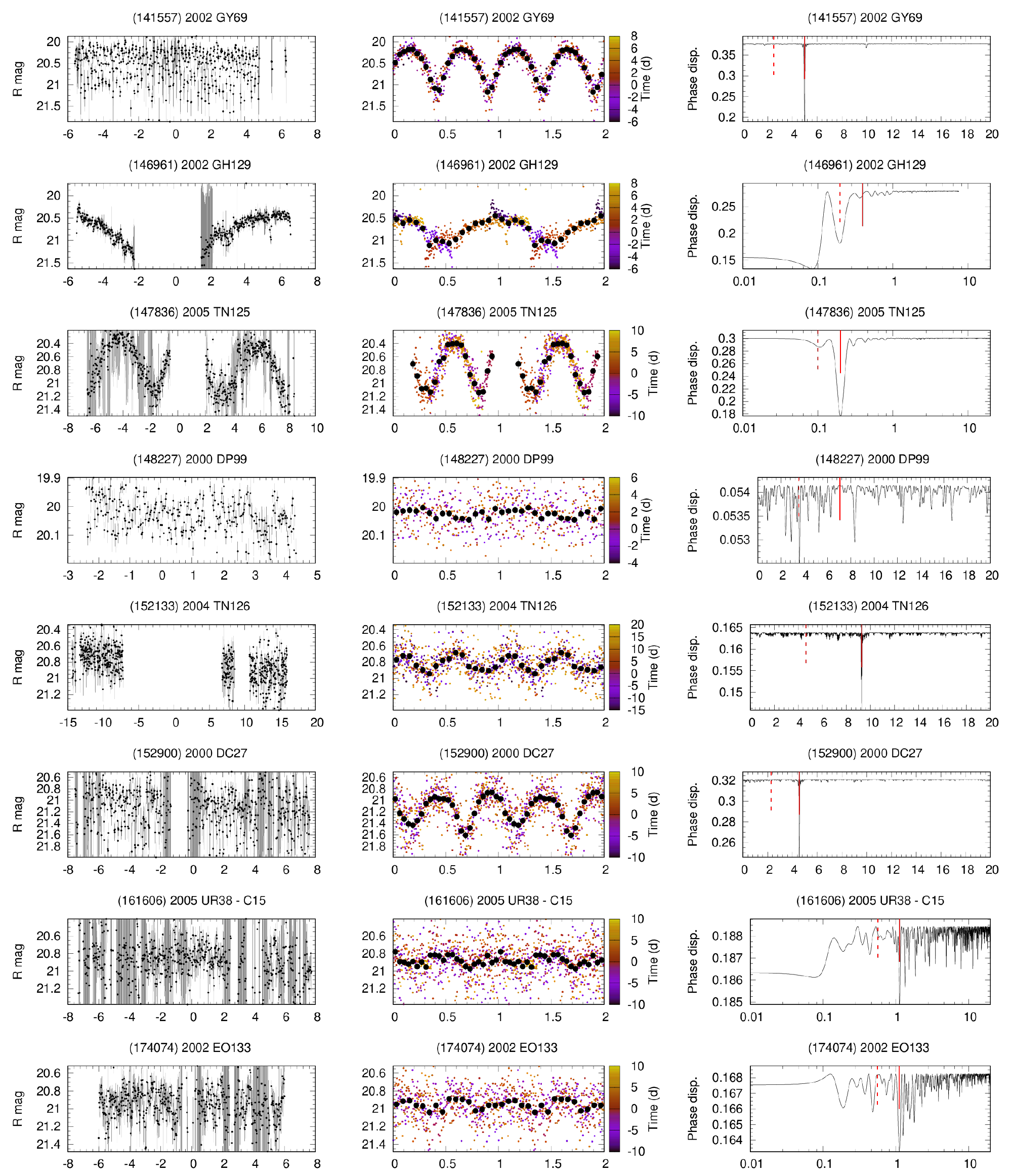}
\figsetgrpnote{Hilda asteroid light curves, continued.}
\figsetgrpend

\figsetgrpstart
\figsetgrpnum{6.11}
\figsetgrptitle{Hilda steroid light curves, continued.}
\figsetplot{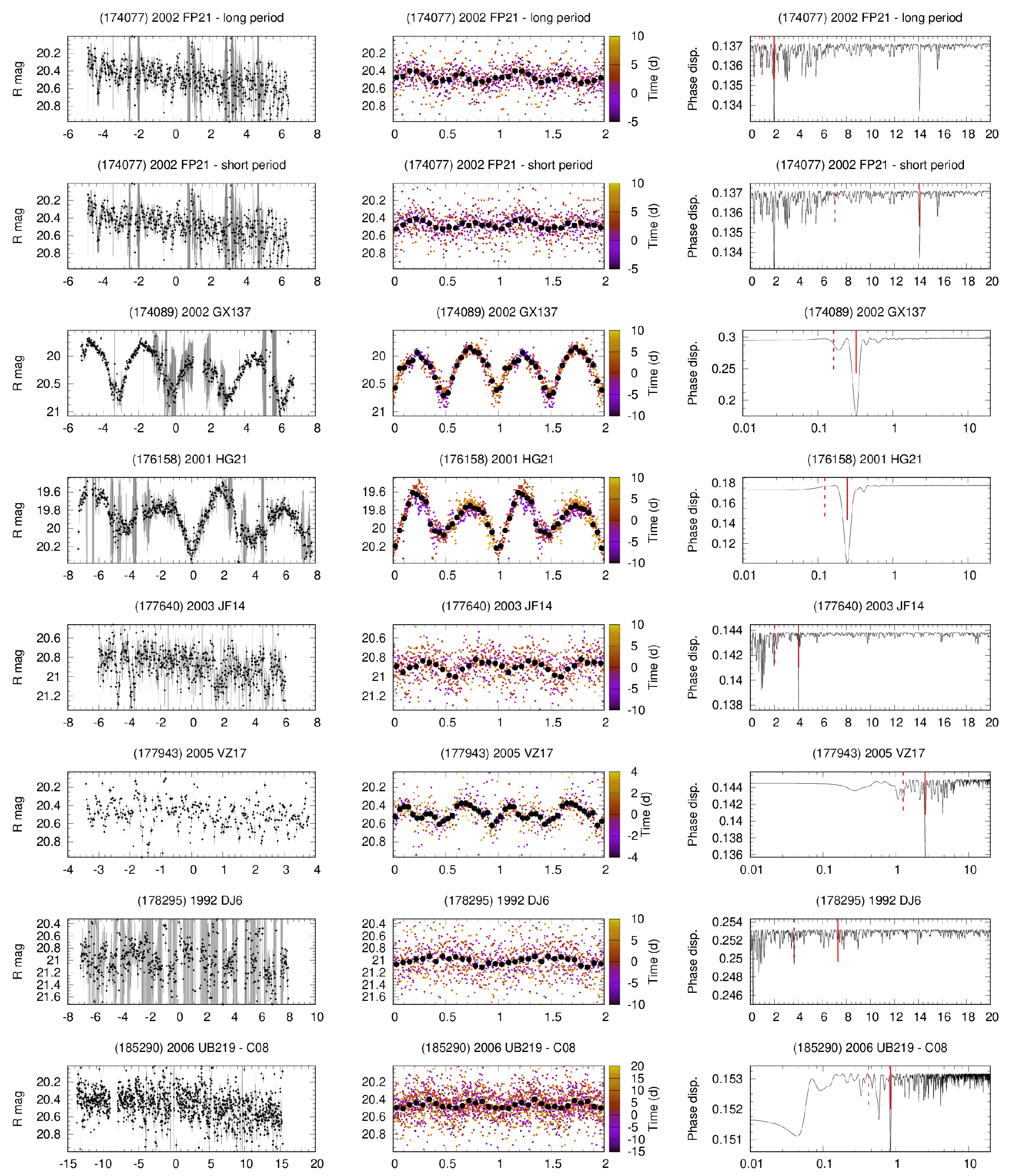}
\figsetgrpnote{Hilda asteroid light curves, continued.}
\figsetgrpend

\figsetgrpstart
\figsetgrpnum{6.12}
\figsetgrptitle{Hilda steroid light curves, continued.}
\figsetplot{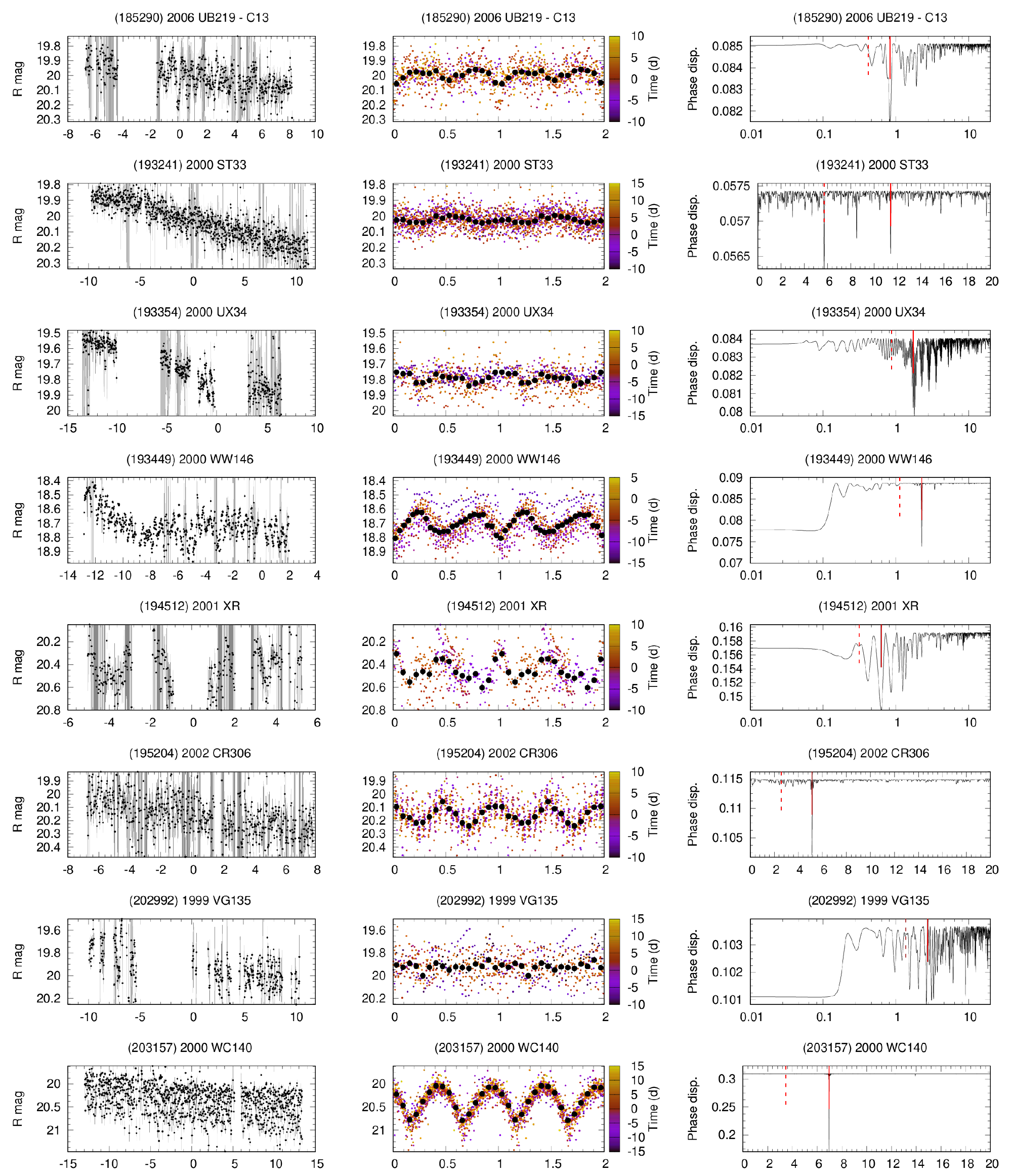}
\figsetgrpnote{Hilda asteroid light curves, continued.}
\figsetgrpend

\figsetgrpstart
\figsetgrpnum{6.13}
\figsetgrptitle{Hilda steroid light curves, continued.}
\figsetplot{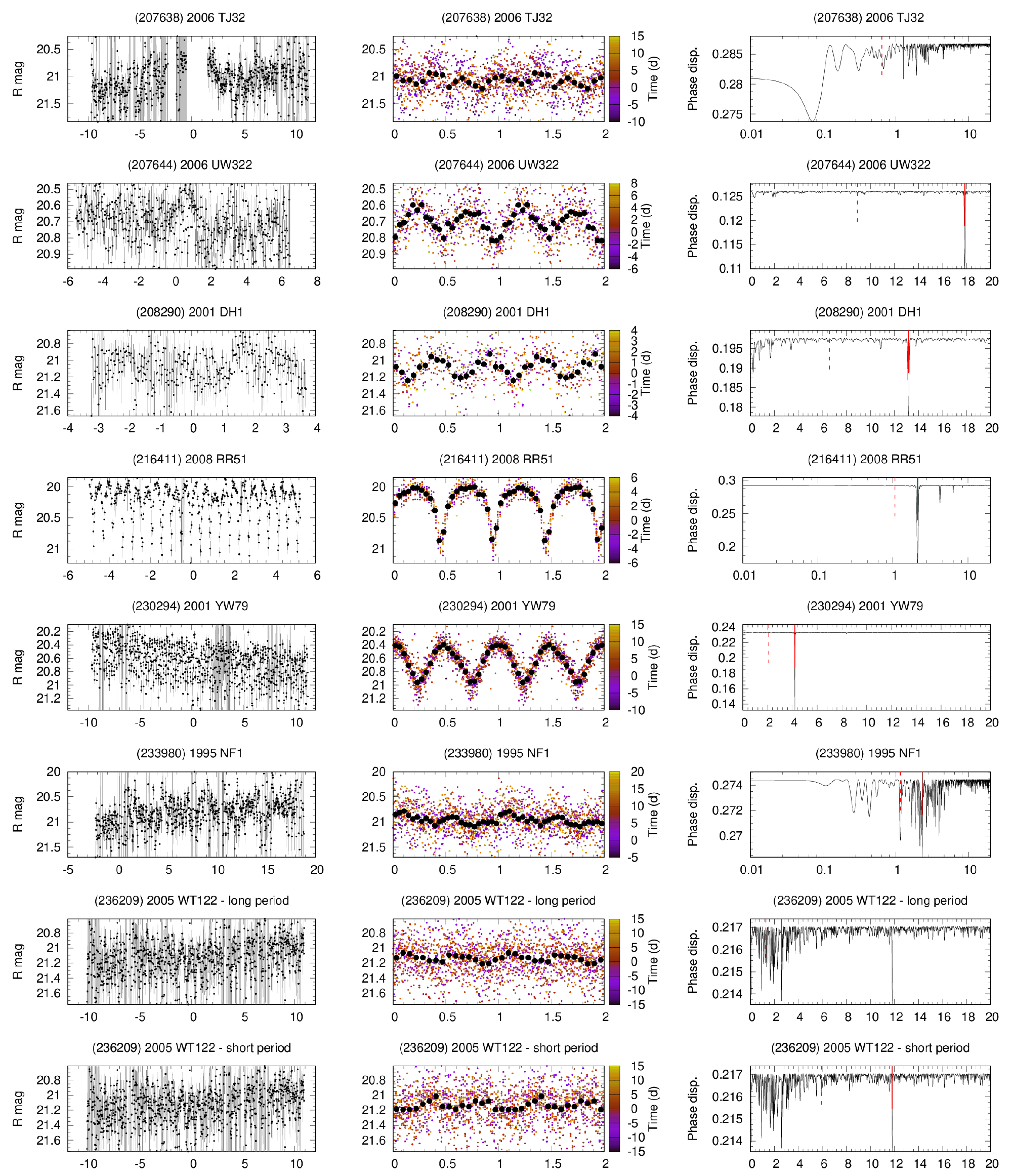}
\figsetgrpnote{Hilda asteroid light curves, continued.}
\figsetgrpend

\figsetgrpstart
\figsetgrpnum{6.14}
\figsetgrptitle{Hilda steroid light curves, continued.}
\figsetplot{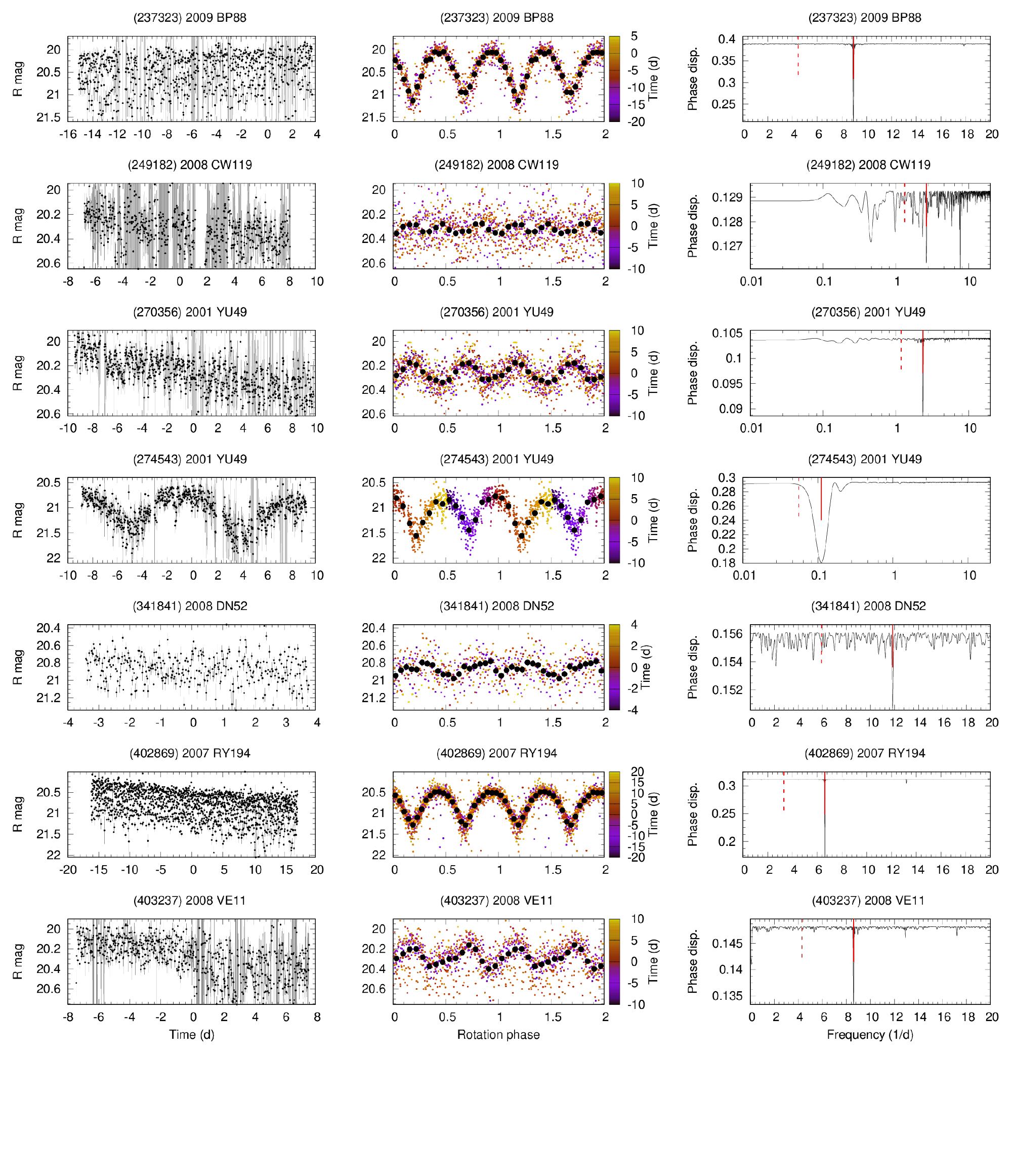}
\figsetgrpnote{Hilda asteroid light curves, continued.}
\figsetgrpend

\figsetend

\begin{figure*}
 \figurenum{6}
    \centering
    \includegraphics[width=\textwidth]{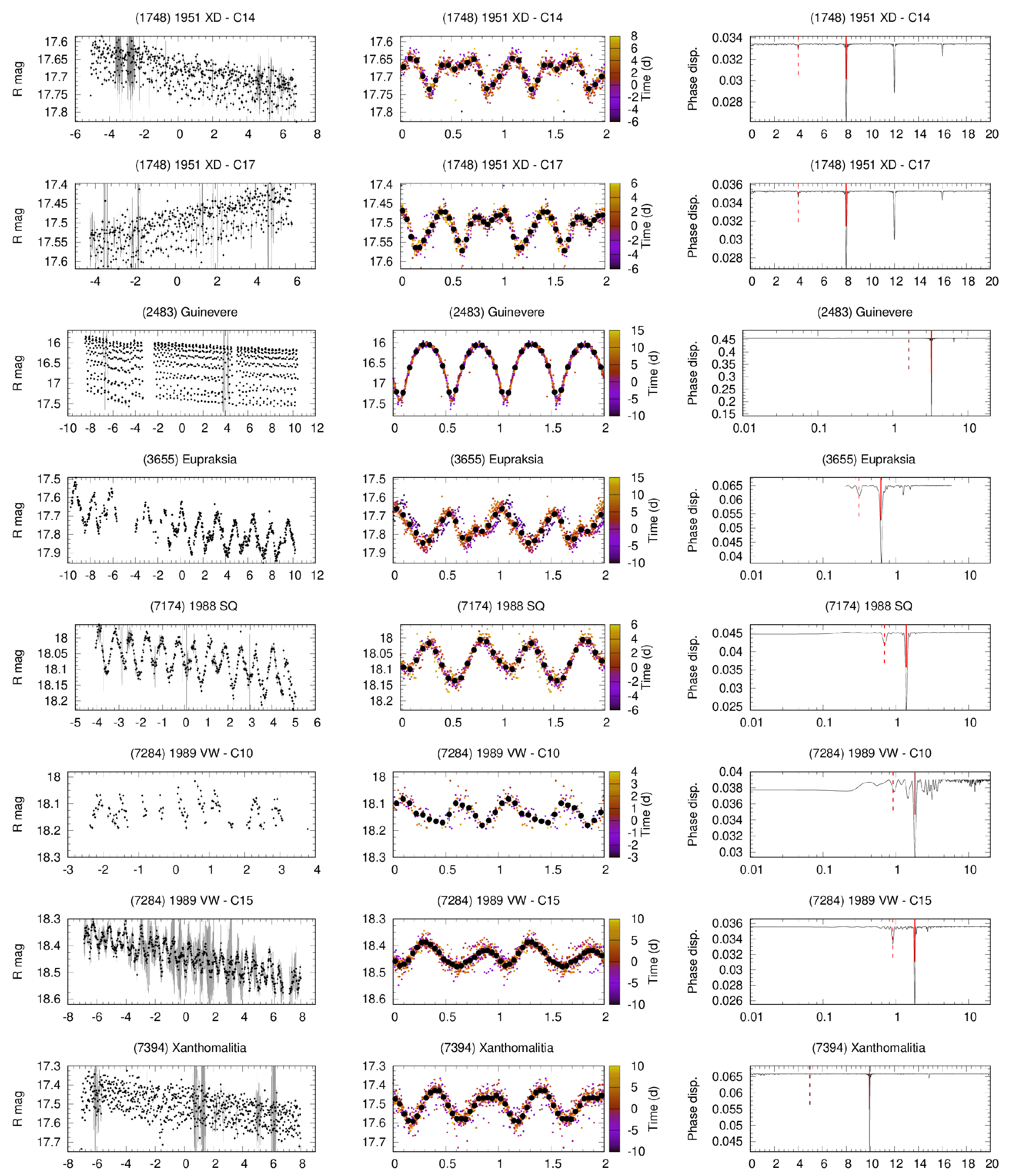}
    \caption{Light curves of Hildas observed by K2. Left: raw light curves. Middle: rectified and folded phase curves. Dots are phase-binned points, color shows time. Right: residual dispersion frequency spectra. Red solid and dashed lines mark frequencies for the single- and double-peak solutions.}
    \label{fig:hildaplot0}
\end{figure*}

\begin{figure*}
 \figurenum{6}
    \centering
    \includegraphics[width=\textwidth]{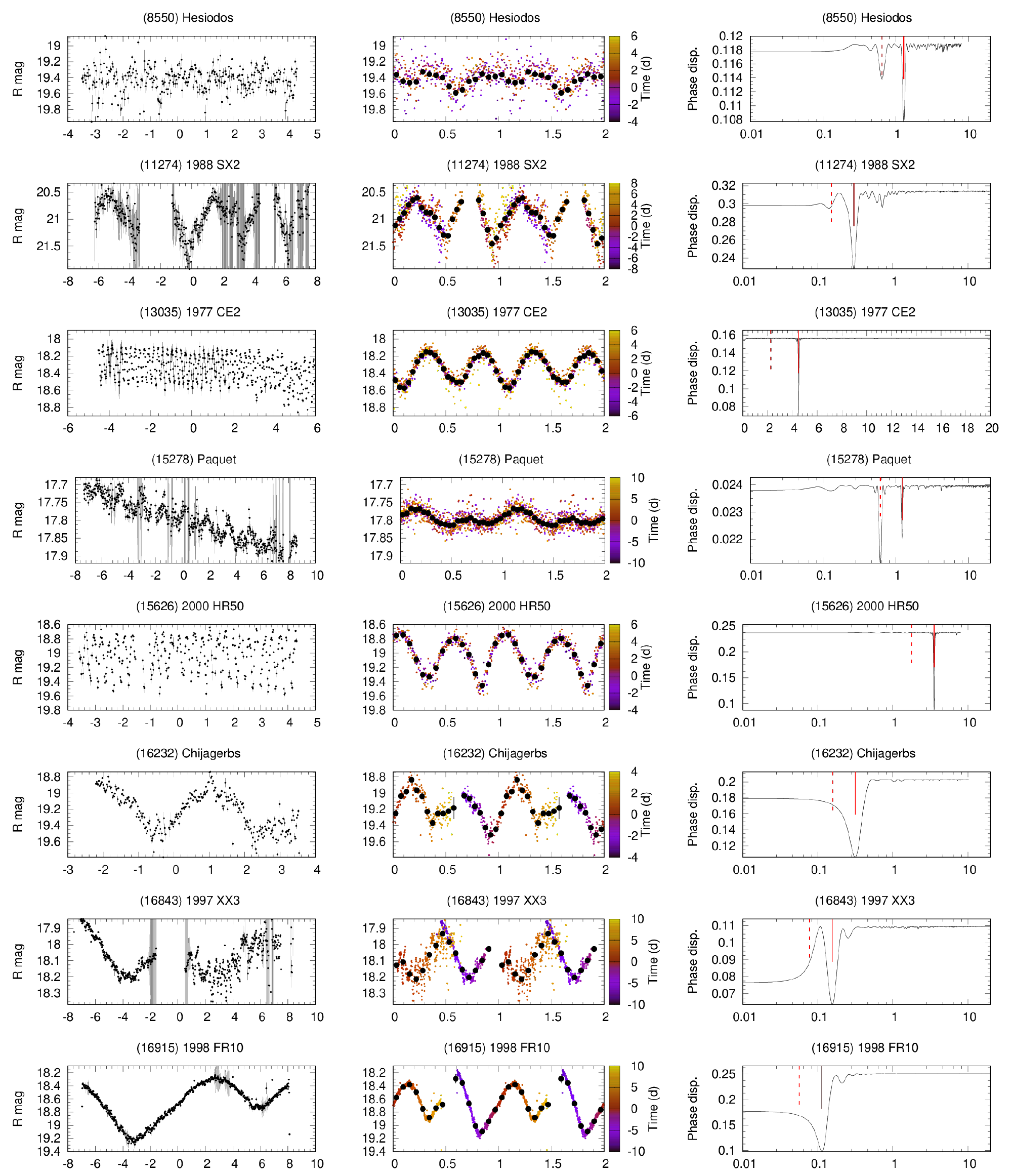}
    \caption{Light curves of Hildas observed by K2. Left: raw light curves. Middle: rectified and folded phase curves. Dots are phase-binned points, color shows time. Right: residual dispersion frequency spectra. Red solid and dashed lines mark frequencies for the single- and double-peak solutions.}
    \label{fig:hildaplot1}
\end{figure*}

\begin{figure*}
 \figurenum{6}
    \centering
    \includegraphics[width=\textwidth]{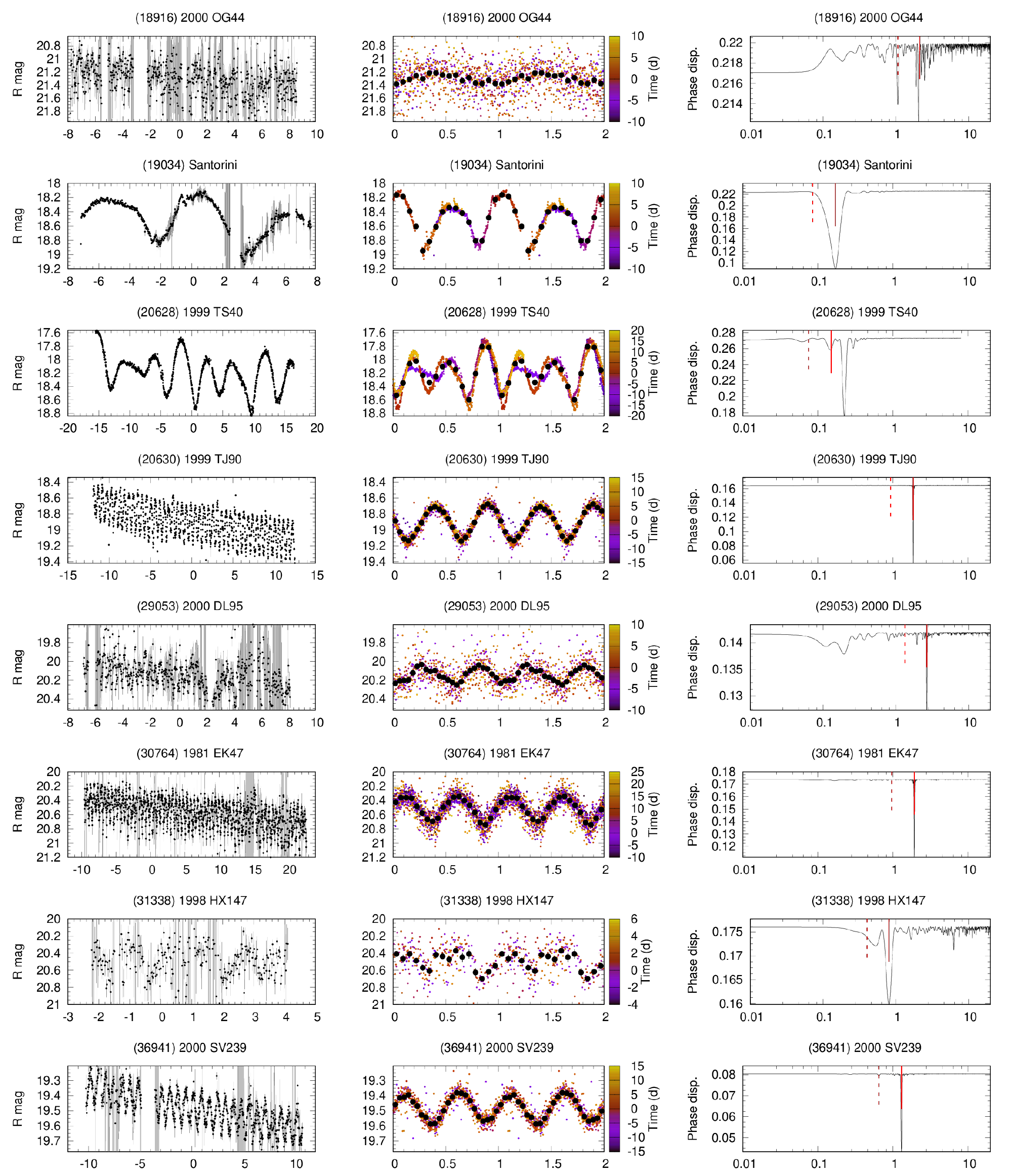}
    \caption{Light curves of Hildas observed by K2. Left: raw light curves. Middle: rectified and folded phase curves. Dots are phase-binned points, color shows time. Right: residual dispersion frequency spectra. Red solid and dashed lines mark frequencies for the single- and double-peak solutions.}
    \label{fig:hildaplot2}
\end{figure*}

\begin{figure*}
 \figurenum{6}
    \centering
    \includegraphics[width=\textwidth]{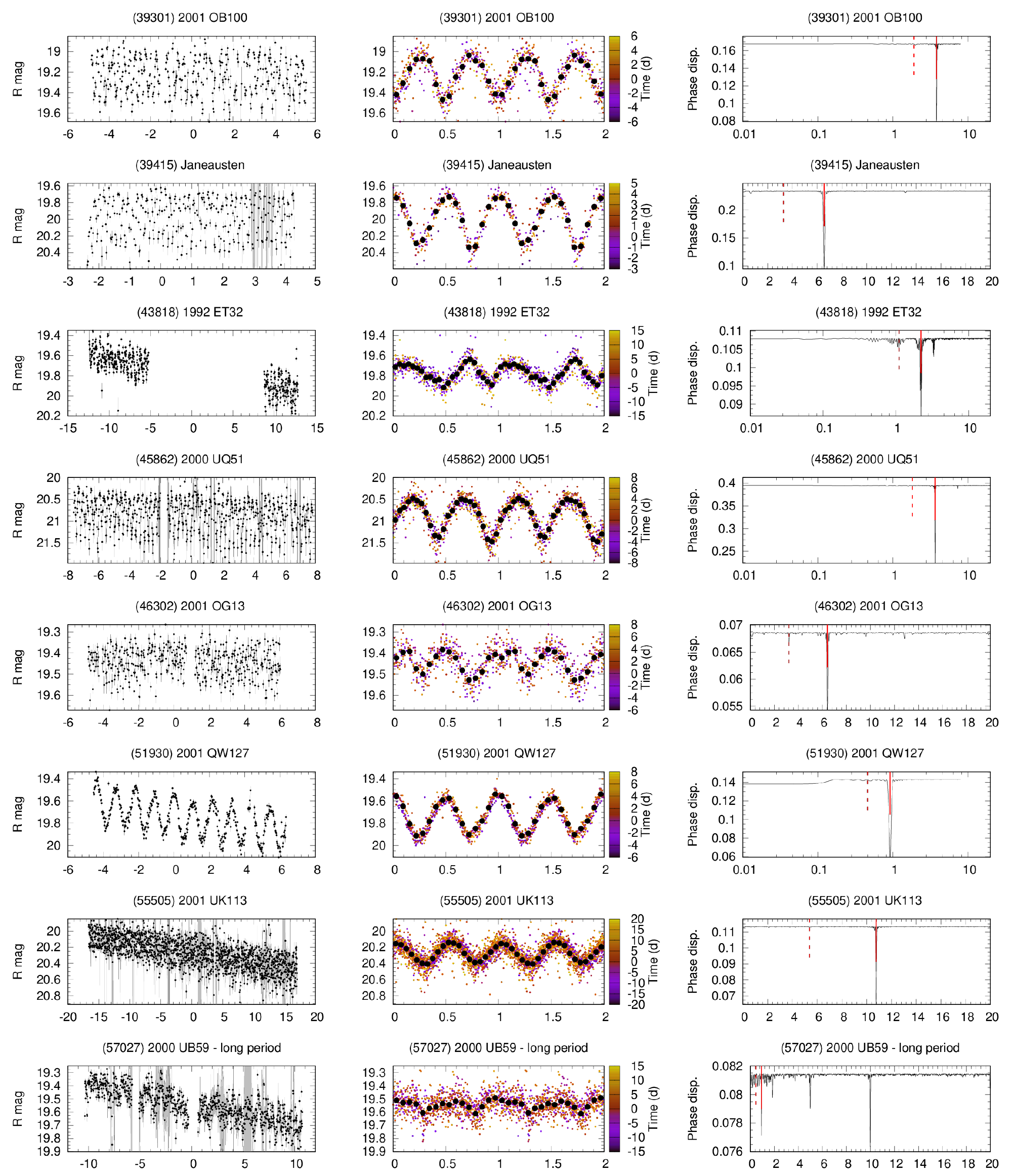}
    \caption{Light curves of Hildas observed by K2. Left: raw light curves. Middle: rectified and folded phase curves. Dots are phase-binned points, color shows time. Right: residual dispersion frequency spectra. Red solid and dashed lines mark frequencies for the single- and double-peak solutions.}
    \label{fig:hildaplot3}
\end{figure*}

\begin{figure*}
 \figurenum{6}
    \centering
    \includegraphics[width=\textwidth]{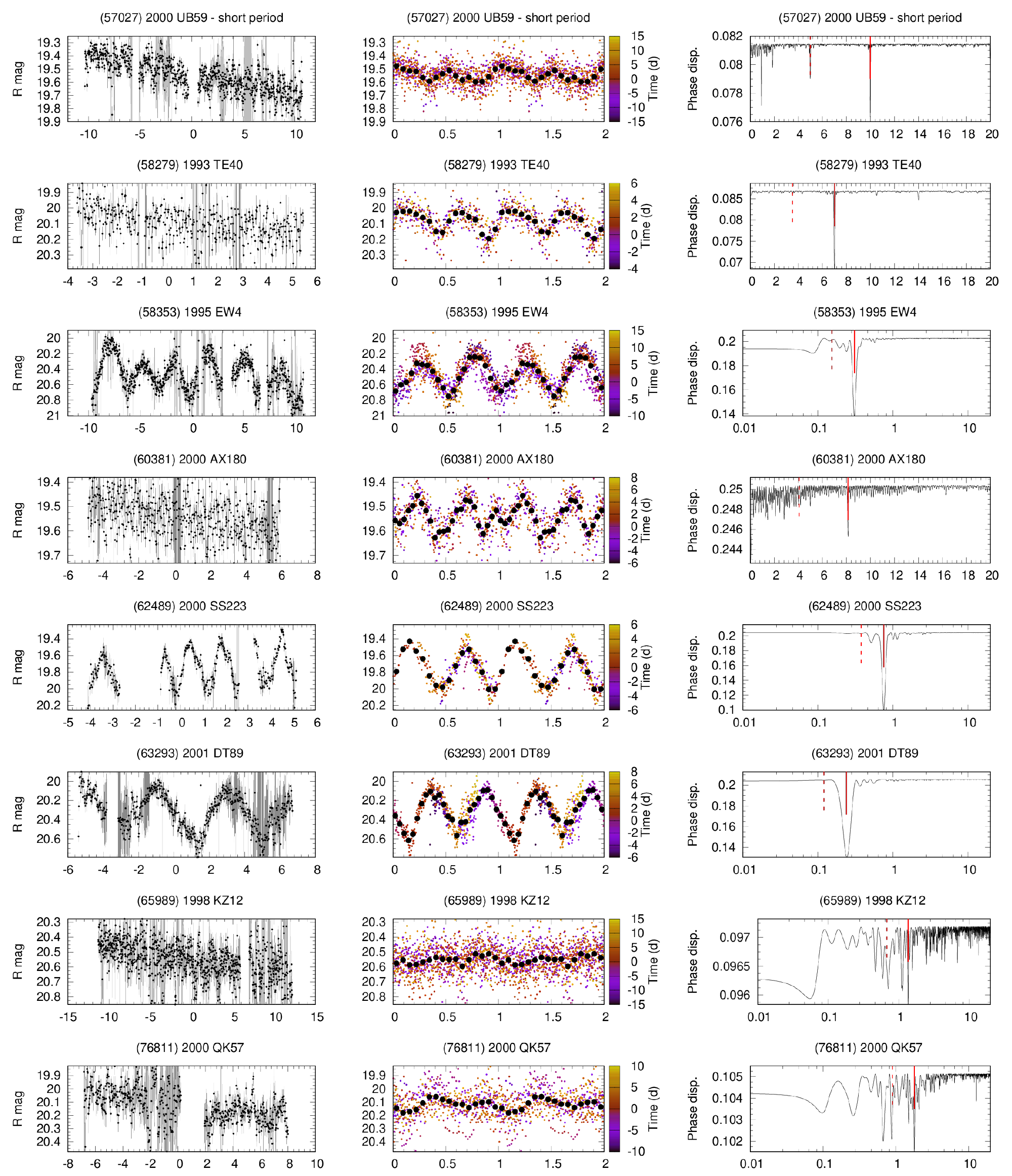}
    \caption{Light curves of Hildas observed by K2. Left: raw light curves. Middle: rectified and folded phase curves. Dots are phase-binned points, color shows time. Right: residual dispersion frequency spectra. Red solid and dashed lines mark frequencies for the single- and double-peak solutions.}
    \label{fig:hildaplot4}
\end{figure*}

\begin{figure*}
 \figurenum{6}
    \centering
    \includegraphics[width=\textwidth]{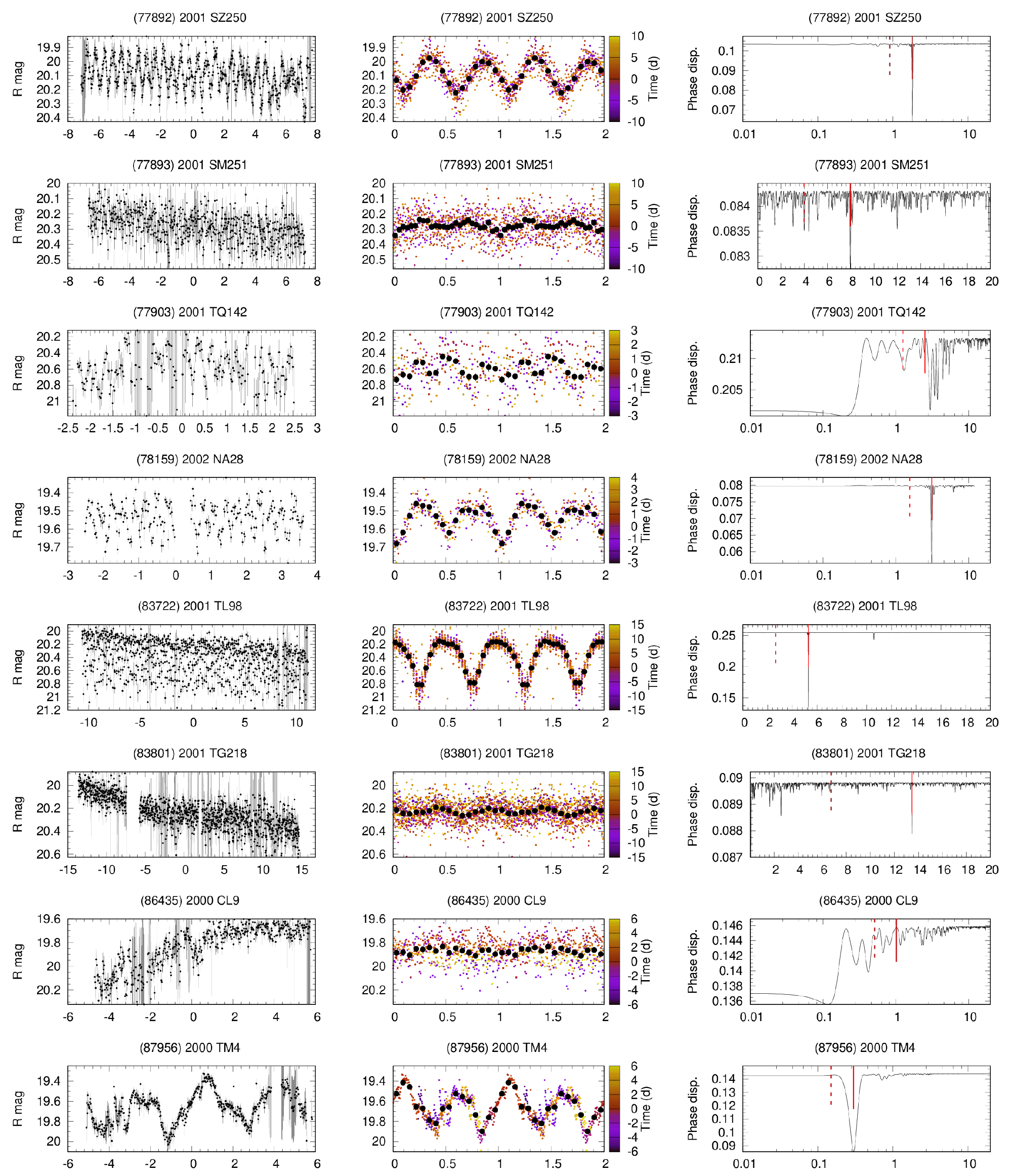}
    \caption{Light curves of Hildas observed by K2. Left: raw light curves. Middle: rectified and folded phase curves. Dots are phase-binned points, color shows time. Right: residual dispersion frequency spectra. Red solid and dashed lines mark frequencies for the single- and double-peak solutions.}
    \label{fig:hildaplot5}
\end{figure*}

\begin{figure*}
 \figurenum{6}
    \centering
    \includegraphics[width=\textwidth]{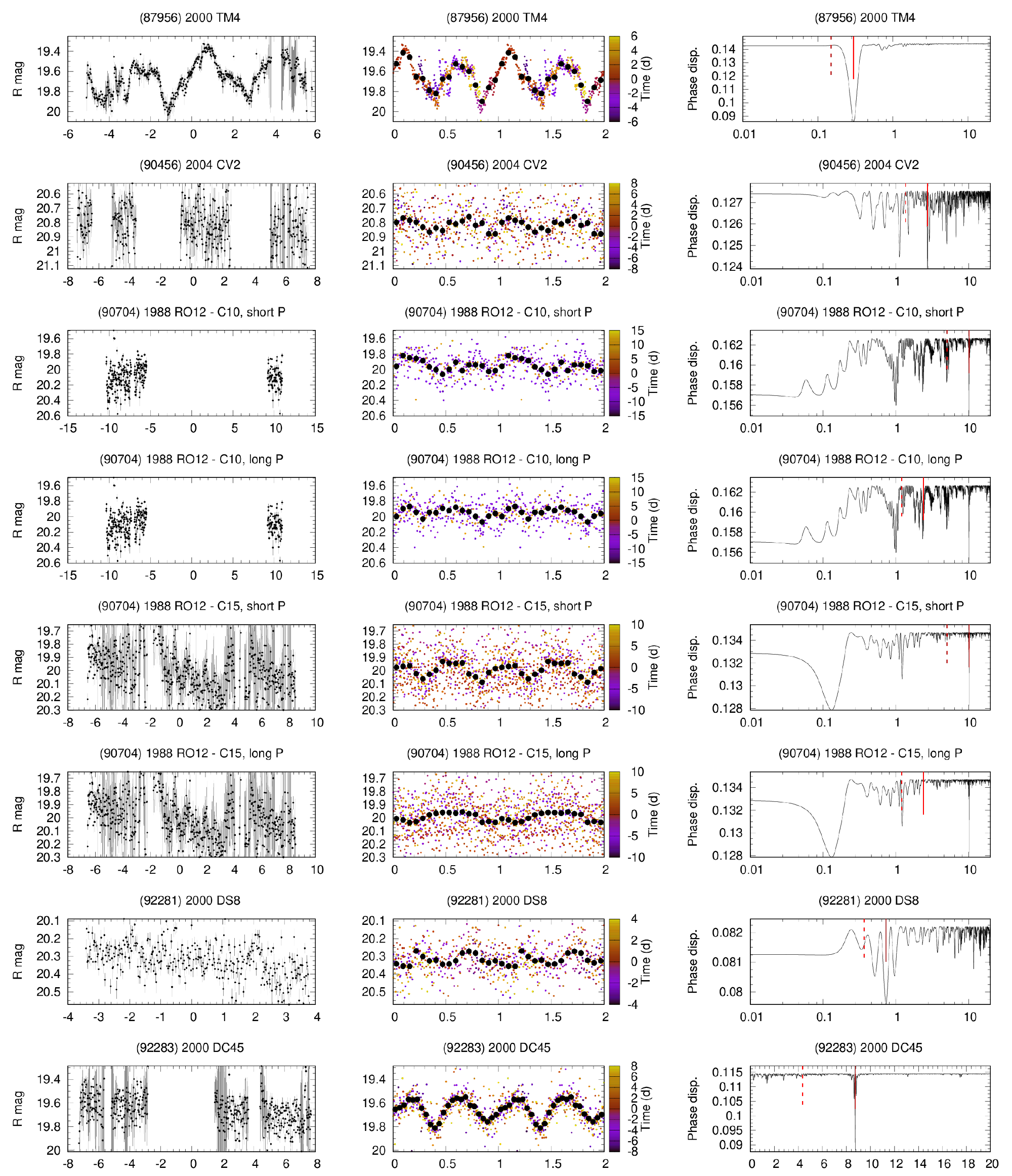}
    \caption{Light curves of Hildas observed by K2. Left: raw light curves. Middle: rectified and folded phase curves. Dots are phase-binned points, color shows time. Right: residual dispersion frequency spectra. Red solid and dashed lines mark frequencies for the single- and double-peak solutions.}
    \label{fig:hildaplot6}
\end{figure*}

\begin{figure*}
 \figurenum{6}
    \centering
    \includegraphics[width=\textwidth]{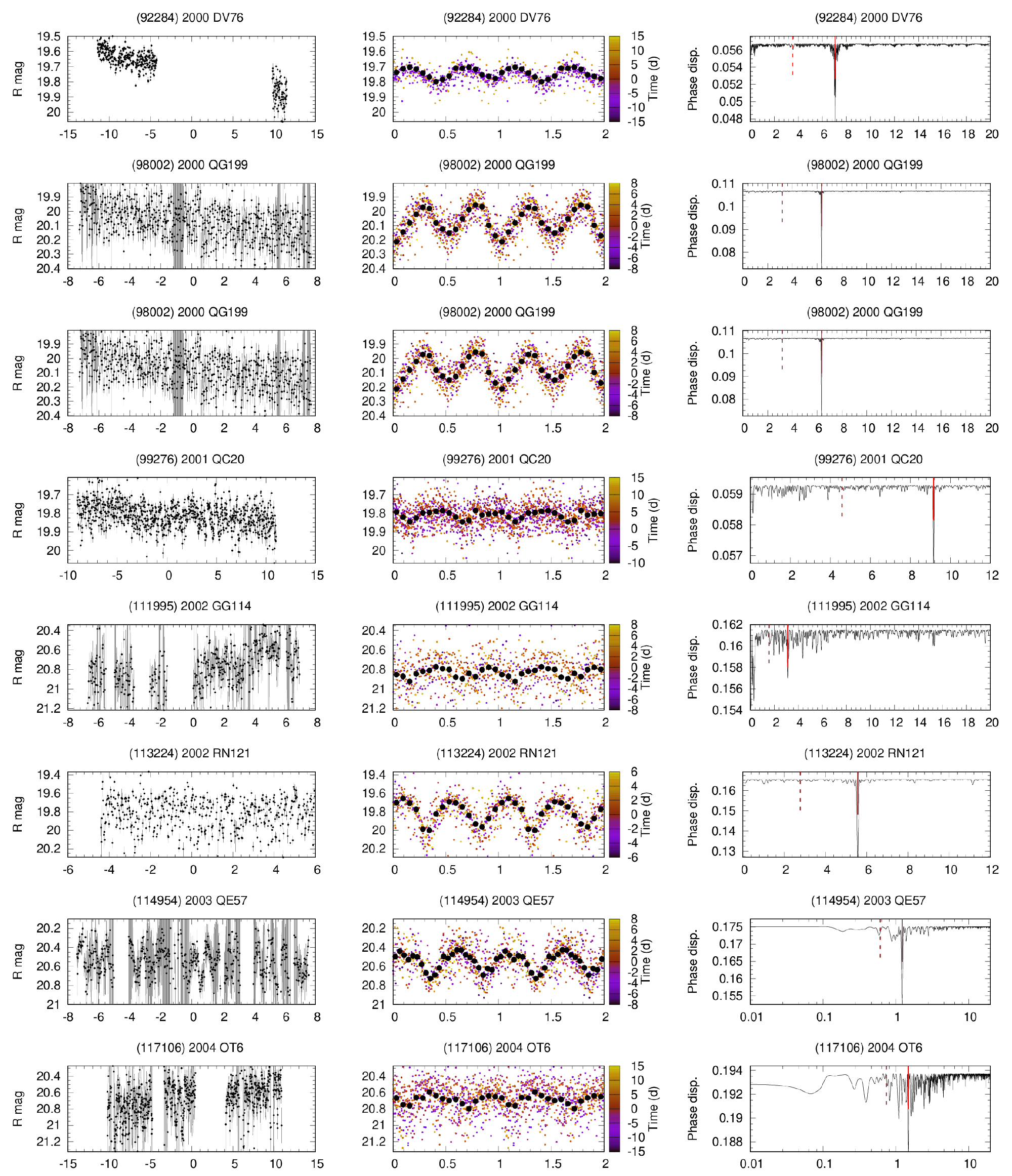}
    \caption{Light curves of Hildas observed by K2. Left: raw light curves. Middle: rectified and folded phase curves. Dots are phase-binned points, color shows time. Right: residual dispersion frequency spectra. Red solid and dashed lines mark frequencies for the single- and double-peak solutions.}
    \label{fig:hildaplot7}
\end{figure*}

\begin{figure*}
 \figurenum{6}
    \centering
    \includegraphics[width=\textwidth]{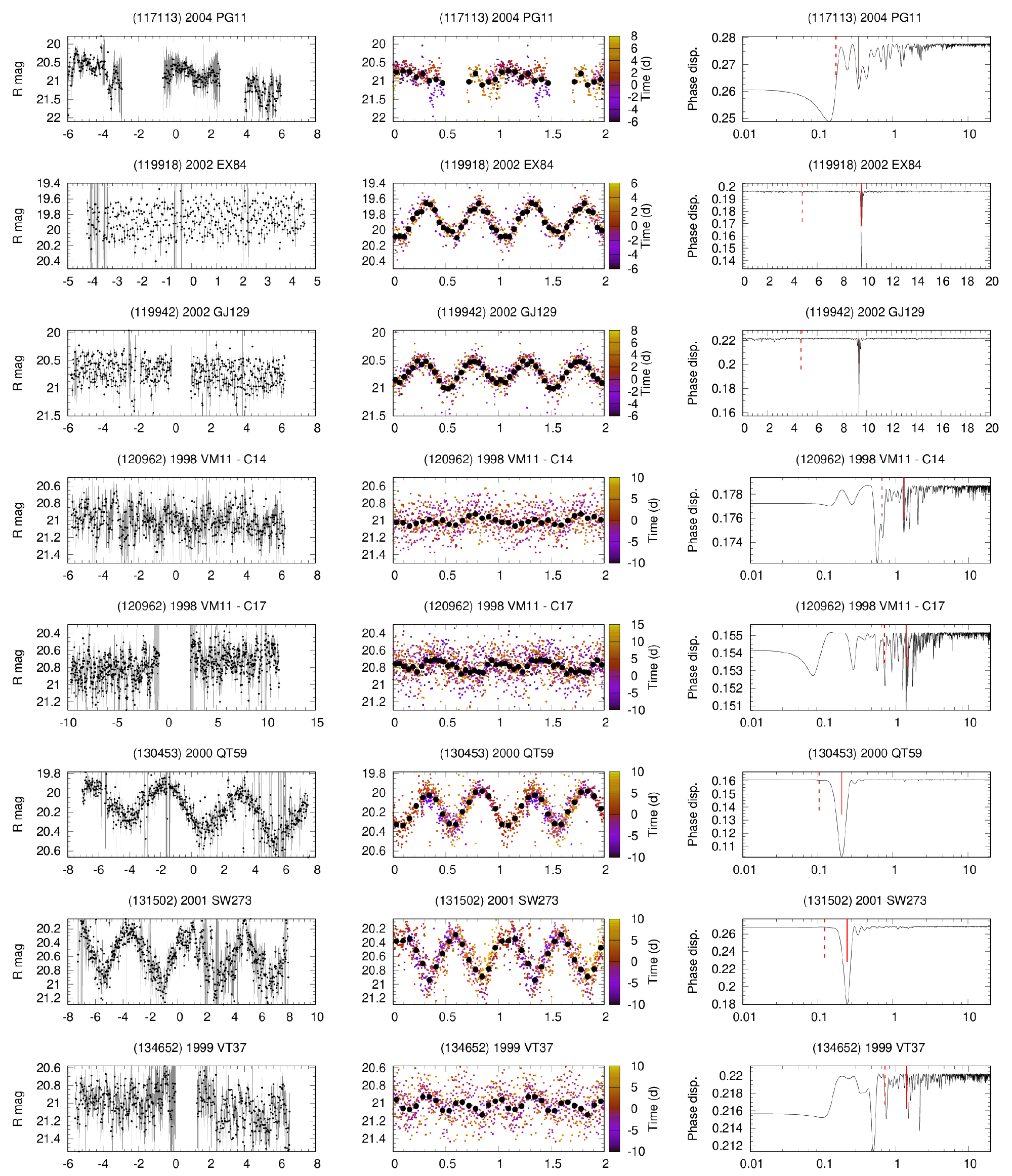}
    \caption{Light curves of Hildas observed by K2. Left: raw light curves. Middle: rectified and folded phase curves. Dots are phase-binned points, color shows time. Right: residual dispersion frequency spectra. Red solid and dashed lines mark frequencies for the single- and double-peak solutions.}
    \label{fig:hildaplot8}
\end{figure*}

\begin{figure*}
 \figurenum{6}
    \centering
    \includegraphics[width=\textwidth]{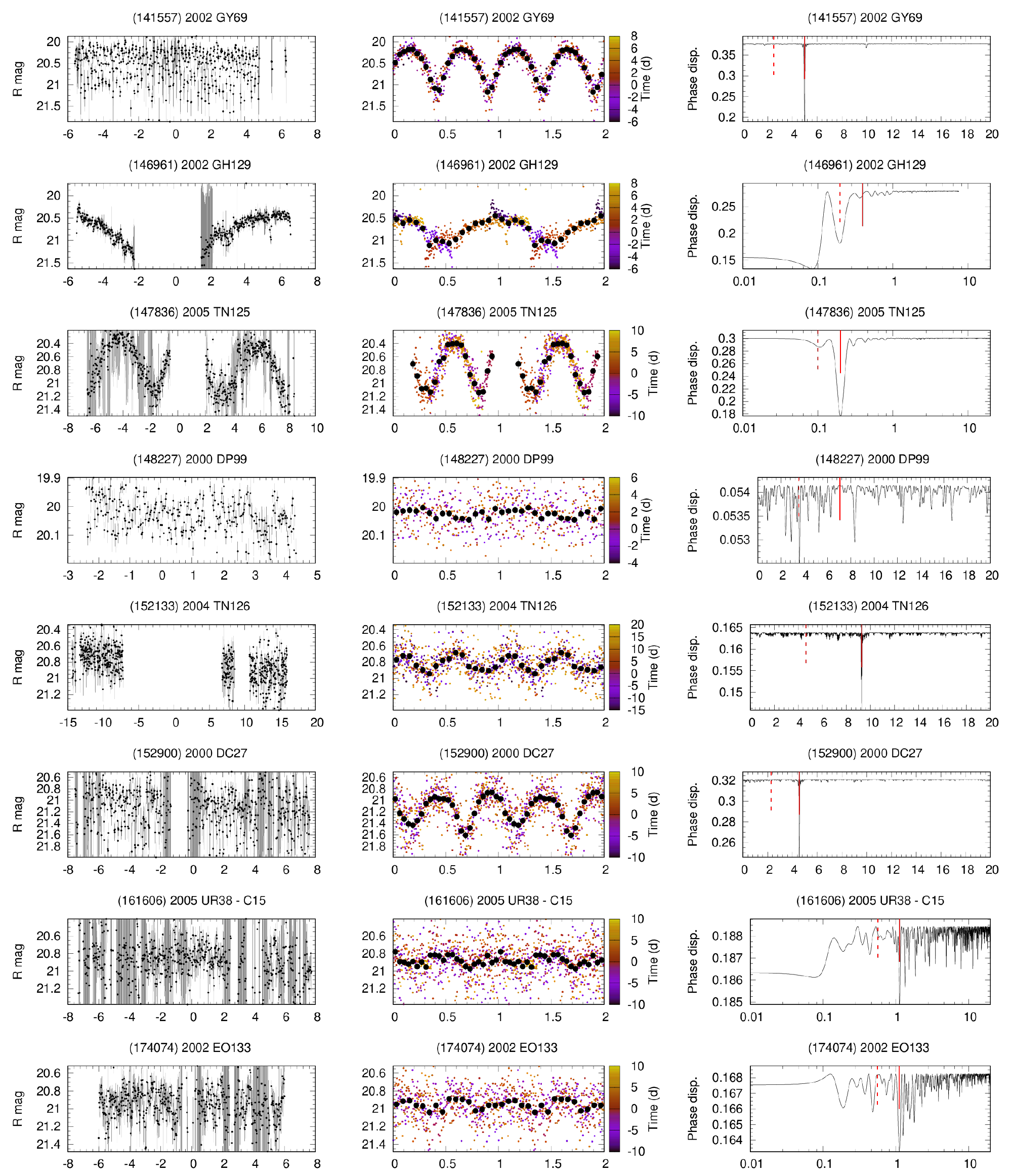}
    \caption{Light curves of Hildas observed by K2. Left: raw light curves. Middle: rectified and folded phase curves. Dots are phase-binned points, color shows time. Right: residual dispersion frequency spectra. Red solid and dashed lines mark frequencies for the single- and double-peak solutions.}
    \label{fig:hildaplot9}
\end{figure*}

\begin{figure*}
 \figurenum{6}
    \centering
    \includegraphics[width=\textwidth]{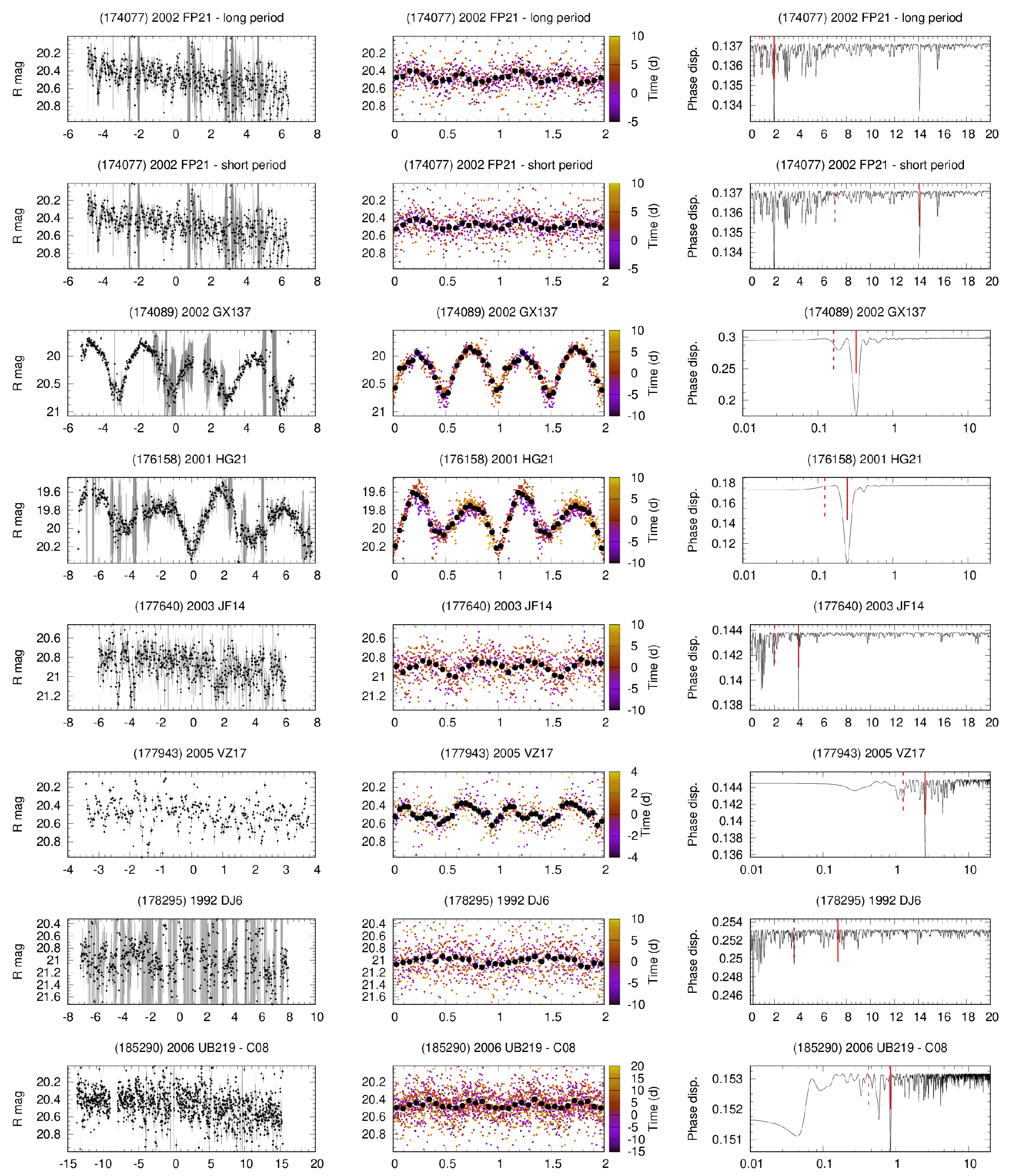}
    \caption{Light curves of Hildas observed by K2. Left: raw light curves. Middle: rectified and folded phase curves. Dots are phase-binned points, color shows time. Right: residual dispersion frequency spectra. Red solid and dashed lines mark frequencies for the single- and double-peak solutions.}
    \label{fig:hildaplot10}
\end{figure*}

\begin{figure*}
 \figurenum{6}
    \centering
    \includegraphics[width=\textwidth]{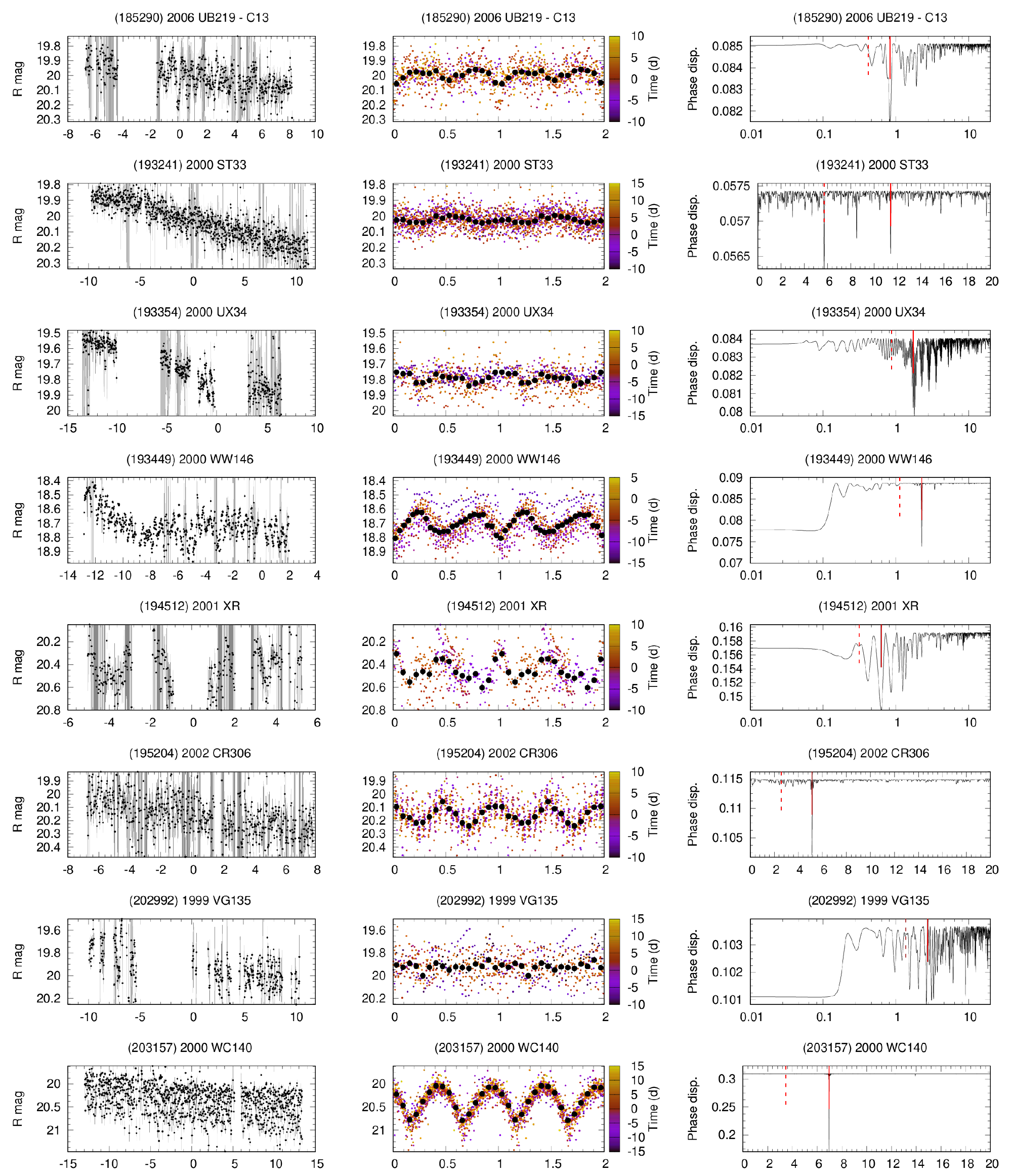}
    \caption{Light curves of Hildas observed by K2. Left: raw light curves. Middle: rectified and folded phase curves. Dots are phase-binned points, color shows time. Right: residual dispersion frequency spectra. Red solid and dashed lines mark frequencies for the single- and double-peak solutions.}
    \label{fig:hildaplot11}
\end{figure*}

\begin{figure*}
 \figurenum{6}
    \centering
    \includegraphics[width=\textwidth]{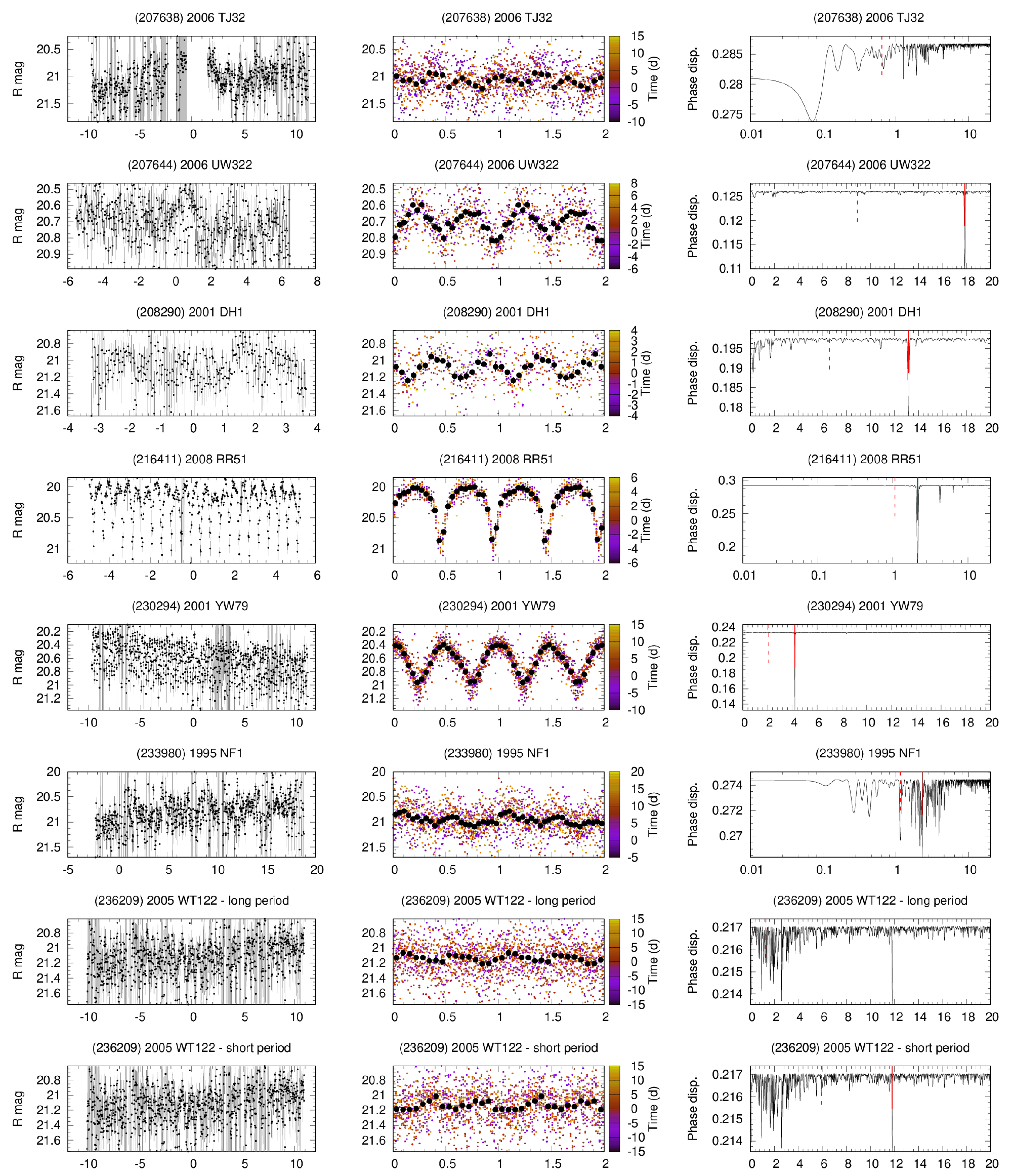}
    \caption{Light curves of Hildas observed by K2. Left: raw light curves. Middle: rectified and folded phase curves. Dots are phase-binned points, color shows time. Right: residual dispersion frequency spectra. Red solid and dashed lines mark frequencies for the single- and double-peak solutions.}
    \label{fig:hildaplot12}
\end{figure*}

\begin{figure*}
 \figurenum{6}
    \centering
    \includegraphics[width=\textwidth]{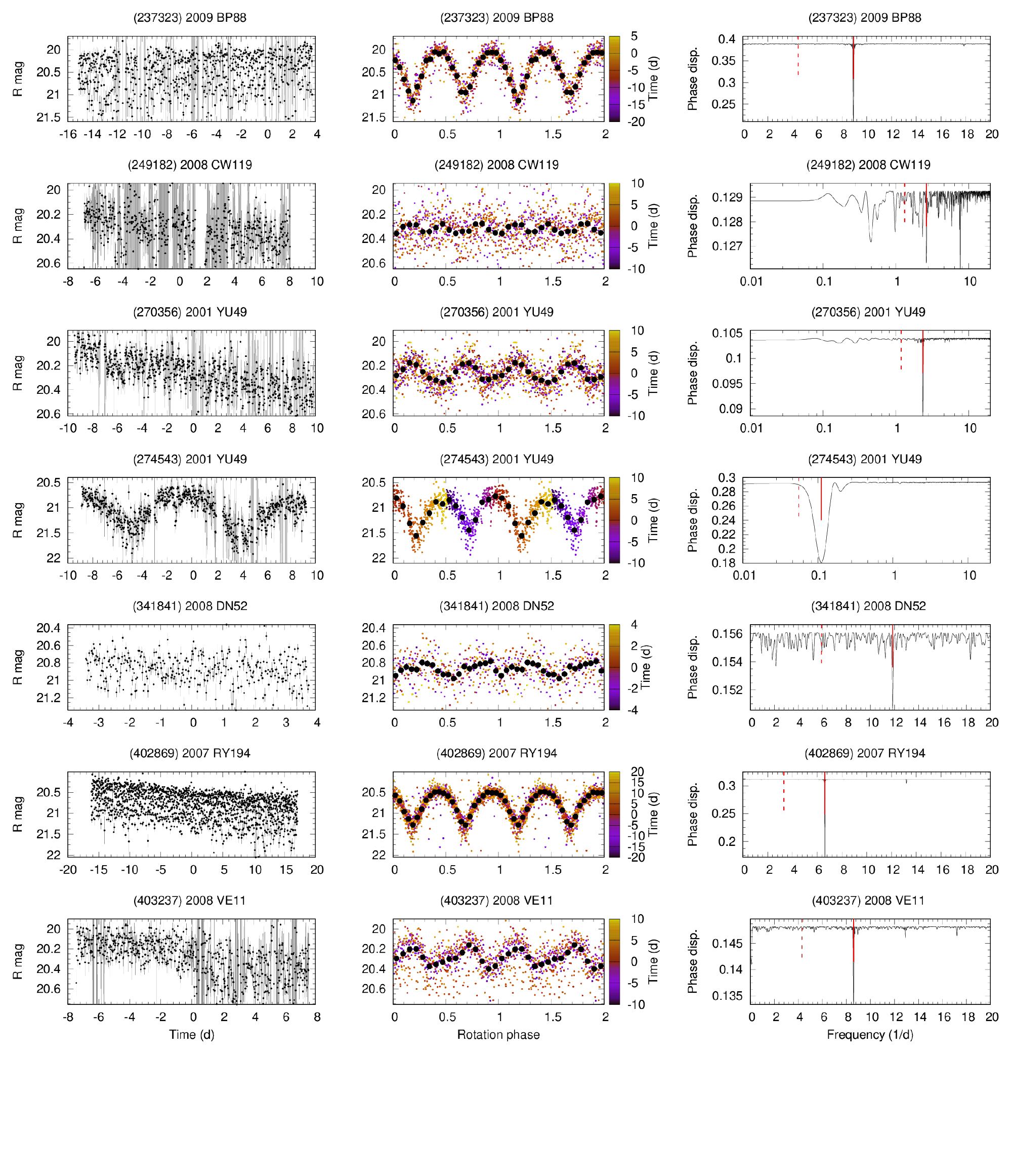}
    \caption{Light curves of Hildas observed by K2. Left: raw light curves. Middle: rectified and folded phase curves. Dots are phase-binned points, color shows time. Right: residual dispersion frequency spectra. Red solid and dashed lines mark frequencies for the single- and double-peak solutions.}
    \label{fig:hildaplot13}
\end{figure*}

\end{appendix}

\end{document}